\documentclass[aps,prx,reprint,superscriptaddress,longbibliography,floatfix,nofootinbib]{revtex4-2}
\usepackage[T1]{fontenc}
\usepackage[utf8]{inputenc}
\usepackage{lmodern}
\usepackage{amsmath,amssymb,amsthm}
\usepackage{graphicx}
\usepackage{booktabs}
\usepackage{microtype}
\usepackage{xcolor}
\usepackage[colorlinks=true,linkcolor=blue!45!black,citecolor=blue!45!black,
            urlcolor=blue!45!black]{hyperref}
\hypersetup{pdftitle={Noise support controls fermionic magic distillation},
            pdfsubject={Fermionic magic-state distillation}}
            \usepackage{orcidlink}
\newcommand{\Tr}{\operatorname{Tr}}
\newcommand{\id}{I}
\newcommand{\CG}{\mathrm{CG}}

\newcommand{\ket}[1]{\lvert #1\rangle}
\newcommand{\bra}[1]{\langle #1\rvert}
\newcommand{\braket}[1]{\langle #1\rangle}
\newcommand{\cD}{\mathcal D}
\newcommand{\cN}{\mathcal N}
\newtheorem{theorem}{Theorem}
\newtheorem{lemma}{Lemma}[section]
\newtheorem{proposition}{Proposition}
\newtheorem{definition}{Definition}
\newtheorem{corollary}{Corollary}
\newtheorem{conjecture}{Conjecture}
\allowdisplaybreaks[2]
\AtBeginDocument{\lefthyphenmin=2\righthyphenmin=3}
\begin{document}
\hypersetup{linkcolor=black,citecolor=blue,urlcolor=black}
\title{Noise support controls fermionic magic distillation}
\author{Dengyao Luo\orcidlink{0009-0004-2383-403X}}
\email{dluo6@ncsu.edu}
\affiliation{Department of Mathematics, North Carolina State University, Raleigh, North Carolina 27695, USA}

\author{Bojko~N.~Bakalov\orcidlink{0000-0003-4630-6120}}
\affiliation{Department of Mathematics, North Carolina State University, Raleigh, North Carolina 27695, USA}
\date{November 30, 2026}

\begin{abstract}
Fermionic magic distillation aims to purify noisy resources using fermionic linear optics. However, target fidelity alone does not determine whether this purification is possible. We show that noise support is an independent determining factor in this problem: sources with the same target fidelity can exhibit opposite distillation behavior. For four-mode fixed-parity sources, we give a complete classification: a state can be distilled to the four-mode Greenberger--Horne--Zeilinger state if and only if it is not convex-Gaussian, and the isotropic threshold is $1/2$. As a no-go result, we prove that no full-rank non-convex-Gaussian source can be purified to any fixed pure non-Gaussian target. For symmetric poisoning of the four-mode target, we determine
the optimal fidelity for any number of input copies. We also consider a mixture of the target and isotropic odd-parity noise as an example between the fixed-parity and full-rank cases. We prove that, for isotropic odd-parity noise with input fidelity $p$, pairing networks followed by arbitrary four-mode fermionic linear optics have optimal fidelity $\max\{1/2,p\}$. These results show that noise support can separate fermionic magic sources into distillable and bound classes even when the initial fidelity is arbitrarily close to one.
\end{abstract}
\maketitle

\setlength{\abovedisplayskip}{8pt plus 2pt minus 2pt}
\setlength{\belowdisplayskip}{8pt plus 2pt minus 2pt}
\setlength{\dbltextfloatsep}{16pt plus 2pt minus 2pt}
\setlength{\textfloatsep}{16pt plus 2pt minus 2pt}

\makeatletter
\@booleantrue\raggedcolumn@sw
\makeatother
\section{Introduction}
\label{sec:introduction}

Magic is an important resource for fault-tolerant quantum computation, and resource distillation is a basic way to obtain high-quality magic resources. A typical example is Clifford magic-state distillation \cite{Bravyi2005Universal}. Previous studies have established general limitations on resource purification
\cite{Fang2020NoGo,Fang2022NoGo,Regula2021Limitations,Regula2022Probabilistic}
and investigated bound magic and undistillability \cite{Campbell2010Bound,Veitch2012Negative,Kitzinger2026Bound}.
The limitations of Gaussian operations in bosonic entanglement distillation
\cite{Eisert2002Distilling,Fiurasek2002Gaussian,Giedke2002Characterization}
and in bosonic and fermionic error correction have also been studied
\cite{Niset2009NoGo,Tang2026ErrorCorrection}.
In the fermionic and Majorana settings studied here, the free operations and 
resources corresponding to Clifford operations and magic resources are 
fermionic linear optics (FLO) and non-Gaussian ancillas, respectively 
\cite{Nayak2008NonAbelian,Bravyi2006Universal,Hebenstreit2019Magic}. A 
natural question is therefore: given a noisy non-Gaussian source, under what 
conditions can FLO purify it into a resource close to a pure magic state?

In this problem, the four-mode Greenberger--Horne--Zeilinger (GHZ) state $\ket M=
(\ket{0000}+\ket{1111})/\sqrt2$ is a widely studied target resource.  
Previous work on this resource has mainly focused on noise restricted in a 
fixed parity 
sector \cite{Bravyi2006Universal,Oszmaniec2014Simulation,Cwiklinski2026Purification}. The situation is different in Majorana devices, where quasiparticle 
poisoning can change parity through single-fermion exchange with the 
environment
\cite{Rainis2012Poisoning,Karzig2021Poisoning,Bhattacharyya2026Dynamics}.
Therefore, even at the same target fidelity, we need to compare sources whose 
noise occupies different parity sectors.

\begin{figure*}[t]
 \centering
 \includegraphics[width=\textwidth]{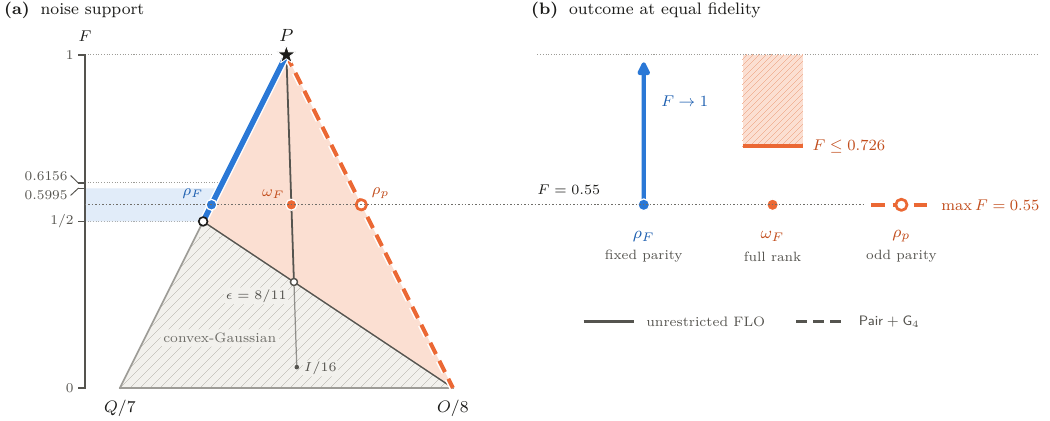}
 \caption{\textbf{Distillability depends on noise support.}
(a) The twirled four-mode slice~\eqref{eq:triangle}, with height
equal to target fidelity.
Gray hatching marks the convex-Gaussian region $a\le b$.
The blue edge contains distillable fixed-parity states;
the orange interior contains bound full-rank states.
The gray line is the white-noise family, entering the free region
at $\epsilon=8/11$.
(b) Comparison at initial fidelity $F=0.55$:
$\rho_F$ can approach the target, but $\omega_F$ has a
fidelity upper bound whose attainability is unknown.
Orange hatching marks the excluded region.
For $\rho_p$ with $p=F$, the exact $\mathsf{Pair+G}_4$ optimum is $F$.
Solid lines denote unrestricted FLO results and dashed lines
results restricted to $\mathsf{Pair+G}_4$.
The unrestricted FLO optimum for $\rho_p$ remains open.}
\label{fig:landscape}
\end{figure*}

We find some sources with the same fidelity but have different distillation behavior under unrestricted FLO (Fig.~\ref{fig:landscape}). An example is  for the same fidelity $F$, when $1/2<F<1$, the fixed-parity source $\rho_F$ can be distilled close to $\ket M$, but the white-noise source $\omega_F$ defined in Sec.~\ref{sec:setting} cannot. 

We describe this phenomenon from two sides. On the positive side, we prove that a fixed-parity four-mode state can be distilled to $\ket M$ if and only if it is not convex-Gaussian (Theorem~\ref{thm:fixed-parity}). Thus, we obtain the isotropic threshold $F_c=1/2$, resolving the interval $1/2<F<0.5995$ left out by previous protocols \cite{Cwiklinski2026Purification}. Our approach is similar to two-qubit entanglement distillation \cite{Horodecki1997Inseparable,Bennett1996Purification,Deutsch1996Privacy}: a filter first raises the fidelity above one half, and a two-copy recurrence then takes it towards one.

On the other hand, we find that, for a physical source of full rank on the whole Fock space and any fixed pure non-Gaussian target, the output fidelity has a uniform upper bound strictly below one (Theorem~\ref{thm:fullrank}(a)). This bound is independent of the number of input copies and is uniform for all positive success probabilities. For the four-mode target $\ket M$,
Theorem~\ref{thm:fullrank}(b) further gives a tighter fidelity upper bound.
Applying it to sources of white-noise gives a lower bound on the positive error proportional to the input infidelity (Corollary~\ref{cor:noise-models}(a)).
This full-rank obstruction reflects a difference between FLO and Clifford magic-state distillation; the latter allows certain full-rank depolarized magic states to be distilled to arbitrary accuracy \cite{Bravyi2005Universal}.\footnote{Reference~\cite{Kitzinger2026Bound} studies undistillability under stabilizer operations with respect to pure nonstabilizer targets. Our target $\ket M$ is itself a qubit stabilizer state in the occupation basis and is a resource relative to FLO.}

The preceding two results treat fixed-parity sources and sources of full rank
on the whole Fock space, respectively.
We are also interested in the case between them, where noise can change parity
but the source support does not yet cover the whole space.
We study $\rho_p=pP+(1-p)O/8$, where $0<p<1$ (Sec.~\ref{sec:setting}).
Here, $P=\ket M\bra M$ and $O$ is the projector onto the odd sector,
so $O/8$ denotes uniform noise on that sector.
Since the even component of this source consists only of the target,
an ideal nondestructive total-parity check will select the target with probability~$p$.

We consider pairing networks built from braiding and bilinear measurements
\cite{Ivanov2001NonAbelian,Bonderson2008Measurement},
with arbitrary FLO allowed on the four remaining modes at the end.
For this class of protocols, we prove that for any number of input copies, the optimal fidelity is
$\max\{1/2,p\}$ (Theorem~\ref{thm:matching}).
The braid-and-fusion protocol of Ref.~\cite{Bravyi2006Universal}
belongs to this class (Proposition~\ref{prop:bravyi-containment}).
Above Bravyi's threshold, this protocol can distill $\rho_F$.
For $\rho_{p=F}$ at the same input fidelity, no protocol in
this class can improve on $F$.

For unrestricted FLO, we further conjecture that,
when $0<p<1$, the optimal fidelity of $\rho_p$
remains $\max\{1/2,p\}$ (Conjecture~\ref{conj:rank-nine}).

The paper is organized as follows.
Section~\ref{sec:setting} defines the operations and sources used in this work.
Sections~\ref{sec:protected}--\ref{sec:matching} present the results and their explanations.
Section~\ref{sec:discussion} discusses quasiparticle poisoning
and supplied parity heralds.
The proofs are given in Appendices~\ref{app:spinor}--\ref{app:geometry}.
\section{Operational setting and noise support}
\label{sec:setting}

Throughout, FLO denotes protocols built from Gaussian unitaries (including those that do not conserve particle number),
independently prepared Gaussian ancillas, occupation-number measurements, classical randomness and feed-forward,
postselection, and discarding modes
\cite{Bravyi2002Fermionic,Terhal2002Simulation,Bravyi2005Lagrangian}.
In the Jordan--Wigner representation, these Gaussian unitaries can be decomposed into nearest-neighbor matchgate circuits
\cite{Knill2001Matchgates,Jozsa2008Matchgates}.

Let $\mathcal F_n$ be the Fock space of $n$ complex fermionic modes.
We use Majorana operators $\gamma_1,\ldots,\gamma_{2n}$ that satisfy
$\{\gamma_j,\gamma_k\}=2\delta_{jk}\id$ and the fermionic annihilation operators $a_j=(\gamma_{2j-1}+i\gamma_{2j})/2$, $j=1,\ldots,n$.
Denote total parity by $\Pi_n=(-1)^{\hat N}$, where $\hat N=\sum_{j=1}^n a_j^\dagger a_j$ is the total occupation-number operator.
Thus, $\Pi_n$ has eigenvalues $+1$ and $-1$ in the even and odd sectors, respectively.
We call an operator \emph{physical} if it commutes with $\Pi_n$
\cite{Banuls2007Entanglement}.
A state has \emph{fixed parity}
if its support lies within a single parity sector.
We take the convex hull $\CG_n$ of all pure Gaussian states as the set of free states
\cite{deMelo2013Power,Oszmaniec2014Simulation}.

The four-mode target state in this work is the GHZ state
\begin{equation}
 \ket M=\frac{\ket{0000}+\ket{1111}}{\sqrt2},
 \quad P=\ket M\bra M.
 \label{eq:target}
\end{equation}
The target $\ket M$ has even parity.
We use $E$ and $O$ to denote the projectors onto the even and odd sectors,
$E=\Pi_+=(I+\Pi_4)/2$ and $O=\Pi_-=(I-\Pi_4)/2$.
$E$ retains the even-parity component of a vector and removes its odd-parity component;
$O$ does the opposite.

Within the even sector, $P$ retains only the target direction.
To distinguish the target from the other directions in this sector, we define $Q=E-P$.
It annihilates $\ket M$ and all odd-parity vectors, while leaving the vectors in the
even subspace orthogonal to $\ket M$ unchanged.
For four modes, the even and odd sectors each have dimension eight.
The orthogonal projectors $P,Q,O$ have ranks $1,7,8$ and satisfy $P+Q+O=I$.
Thus, $Q/7$ and $O/8$ describe uniform target-orthogonal noise in the same
and opposite parity sectors, respectively.

By \emph{noise support relative to $P$}, we mean the support of the
positive error block $(I-P)\rho(I-P)$, namely the target-orthogonal
subspace carrying a nonzero error weight.
For families $\rho=FP+(1-F)\sigma_{\rm err}$ with
$\sigma_{\rm err}P=0$, this is $\operatorname{supp}\sigma_{\rm err}$
when $F<1$. For these families with $0<F<1$, the full source support is
the direct sum of the target direction and the noise support.

Let $\rho$ denote the source.
For $N$ independent copies, let $\Phi$ denote an accepted FLO operation that leaves four output modes.
Then $\Phi(\rho^{\otimes N})$ is the unnormalized output, whose trace is the success probability.
When this probability is nonzero, we denote the normalized output conditioned on success by $\rho_{\rm out}= \Phi(\rho^{\otimes N}) / \Tr\Phi(\rho^{\otimes N}).$
For the resulting output, we define the fidelity $F_M(\rho_{\rm out})=\Tr(P\rho_{\rm out})$ as its weight along the target direction; the closer it is to one, the closer the output is to the distillation target.
Every convex-Gaussian state has fidelity at most $1/2$ with this target.

Taking the supremum over all accepted FLO operations with nonzero success probability gives
\begin{equation}
 f_N(\rho)=
 \sup_{\Phi:\,\Tr\Phi(\rho^{\otimes N})>0}
 \frac{\Tr[P\Phi(\rho^{\otimes N})]}
      {\Tr\Phi(\rho^{\otimes N})}.
 \label{eq:distillation-definition}
\end{equation}

As the number of inputs $N$ is allowed to increase, the condition $\sup_{N\ge1}f_N(\rho)=1$ means that the output can approach $\ket M$ arbitrarily closely.
In this case, we call $\rho$ \emph{distillable} with respect to $\ket M$.
If $\rho$ is non-convex-Gaussian and $\sup_{N\ge1}f_N(\rho)<1$, we call it \emph{bound} with respect to $\ket M$.

To compare the effects of noise support, we use a Gaussian twirl to construct a comparison space.
Here, a Gaussian twirl means randomly applying and averaging over a collection of Gaussian unitaries that leave the target invariant.
A general four-mode state contains information about other directions and coherences in addition to its weights on $P,Q,O$.
Our twirl randomly mixes finitely many Gaussian unitaries that leave $P$ invariant, preserving the target fidelity while averaging out the remaining details.
It therefore maps any physical four-mode state to the following form with three parameters:
\begin{equation}
 \begin{gathered}
 \rho(a,b,c)=aP+\frac b7Q+\frac c8O,\\
 a+b+c=1,\quad a,b,c\ge0,
 \end{gathered}
 \label{eq:triangle}
\end{equation}
where $a$ is the target weight,
$b$ is the total weight in the same-parity error subspace $Q$,
and $c$ is the total weight in the opposite-parity subspace $O$.
Since the twirl preserves target fidelity, $a$ also equals the fidelity of the original input.

The region $a\le b$ is the convex-Gaussian set, which is the free region in this comparison space. The precise criterion is given in Appendix \ref{app:geometry} (see \cite{Oszmaniec2014Simulation}).
When $c=0$, the free region is $a\le1/2$, and there is no noise in the odd sector.

Fixing the same target fidelity $1/2<F<1$, we place the remaining weight $1-F$ in different subspaces to obtain
\begin{align}
 \rho_F&=FP+(1-F)Q/7
 &&\text{(even-parity noise)},\nonumber\\
 \omega_F&=FP+(1-F)(Q+O)/15
 &&\text{(white noise)},
 \label{eq:same-fidelity}\\
 \rho_{p=F}&=FP+(1-F)O/8
 &&\text{(odd-parity noise)}.\nonumber
\end{align}
The source $\rho_F$ places all its noise in $Q$, so the input remains in the even sector.
The noise in $\omega_F$ is uniformly distributed over the fifteen directions orthogonal to the target, so both parity sectors contain noise.
The source $\rho_{p=F}$ places all its errors in $O$, and is not full rank.
The above sources allow us to see how changing the noise support affects distillation in the same fidelity. The construction of this twirl and the Majorana-error interpretations of $Q/7$ and $O/8$
are given in Appendix~\ref{app:geometry}.
\section{Distillation of fixed-parity sources}
\label{sec:protected}

In this section, we study the inputs supported entirely in a single parity sector and answer whether every non-convex-Gaussian source of fixed parity can be distilled to the GHZ state $\ket M$.

\begin{theorem}[Fixed-parity distillation]
\label{thm:fixed-parity}
Let $\rho$ be a four-mode state supported in a single parity sector.
Under FLO,
\begin{equation}
 \sup_{N\ge1}f_N(\rho)=1
 \quad\Longleftrightarrow\quad
 \rho\notin\CG_4.
 \label{eq:fixed-parity-theorem}
\end{equation}
For $\rho\in\CG_4$, the optimum is $f_N(\rho)=1/2$ for every
$N\ge1$.
\end{theorem}

We first consider why a convex-Gaussian input cannot be distilled to $\ket M$.
Tensor products of such states remain convex-Gaussian, and FLO with postselection preserves this property. Thus, regardless of the number of inputs, the output fidelity is at most $1/2$. This bound can be attained by preparing the vacuum.

On the other hand, we construct a distillation procedure for a non-convex-Gaussian input. If $\rho$ lies in the odd sector, FLO can reversibly convert it to the even sector, so it suffices to consider the even sector. Its initial fidelity may be below $1/2$, so the first step is to turn it into a suitable input for the subsequent protocol. We use the four-mode concurrence criterion \cite{Oszmaniec2014Simulation} to construct a Gaussian filter that produces an output with fidelity greater than $1/2$ with a positive success probability. Since the subsequent two-copy protocol is designed for an isotropic family, we apply a Gaussian twirl that preserves target fidelity and distributes the filtered output's weight uniformly over the $Q$ subspace, obtaining
\begin{equation}
 \rho_F=FP+\frac{1-F}{7}Q,\qquad F>1/2.
 \label{eq:even-family}
\end{equation}
Every non-convex-Gaussian fixed-parity input is thereby reduced to the same standard family.

\begin{figure*}[t]
 \centering
 \includegraphics[width=\textwidth]{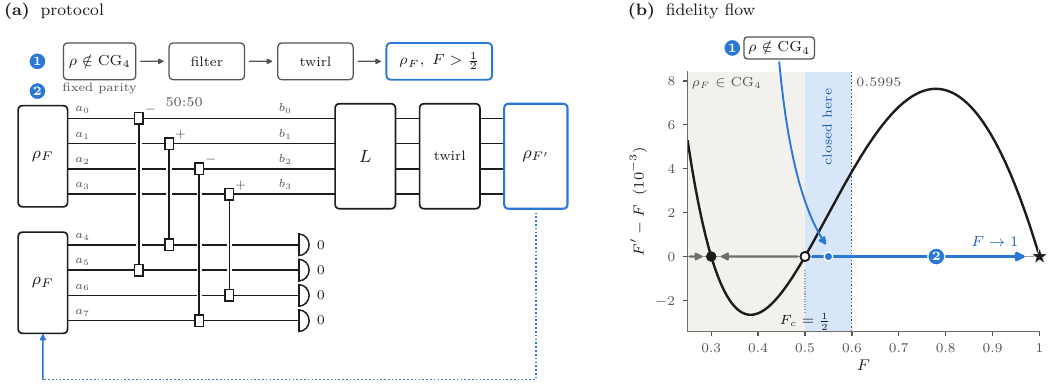}
 \caption{\textbf{A two-copy protocol reaches the free boundary.}
 (a) A Gaussian filter and twirl convert any non-convex-Gaussian
fixed-parity four-mode source into $\rho_F$ with $F>1/2$.
Each round combines two copies using beam splitters, vacuum
postselection, the filter $L=2^{(\widehat N_{\mathrm{out}}-4)/4}$,
and a twirl (Appendix~\ref{app:fixed-parity}).
(b) The resulting fidelity gain $F'-F$
[Eqs.~\eqref{eq:main-recurrence}--\eqref{eq:main-gain}].
Iteration approaches $F=1$ for every $F>1/2$.
Gray shading denotes the convex-Gaussian members of the twirled
family; the curved arrow shows the preprocessing of the general
source in (a).
Blue shading marks $1/2<F<0.5995$, below the sufficient threshold
of Ref.~\cite{Cwiklinski2026Purification}.}
\label{fig:protocol}
\end{figure*}

It remains to increase the fidelity from above $1/2$ towards $1$.
We apply the two-copy FLO protocol in Fig.~\ref{fig:protocol} to this standard family, which is analogous to the recurrence protocols in Ref.~\cite{Bennett1996Purification,Deutsch1996Privacy}. 
Each round processes two identical copies of $\rho_F$ and twirls the accepted output. 
The successful output therefore remains in this standard family and can serve as an 
input to the next round. We track the ratio $r$, between the weight of the target and the weight of a noise direction, in each round to follow the entire iteration.
In $\rho_F$, the target weight is $F$, and the weight of each of the other seven noise directions is $(1-F)/7$, so $r=7F/(1-F)$.
Thus, $F>1/2$ is equivalent to $r>7$.
The protocol acts on $r$ as follows:
\begin{equation}
 r'=\frac{7(2r^2+r+3)}{13r+17}.
 \label{eq:main-recurrence}
\end{equation}
We compare $r'$ with $r$ in each round:
\begin{equation}
 r'-r=\frac{(r-7)(r-3)}{13r+17}.
 \label{eq:main-gain}
\end{equation}
For $r>7$, the right-hand side is positive, so the ratio increases strictly in every round. If this increasing sequence converged to a finite value, continuity of the recurrence would require that value to be a fixed point.
However, the only finite fixed points are $3$ and $7$. We must have $r\to\infty$, and the output fidelity tends to 1.

Thus, $\rho_F$ is distillable whenever $F>1/2$, and its threshold lies at the free boundary $F_c=1/2$.
This answers the FLO distillability question on the interval $1/2<F<0.5995$ left open by Ref.~\cite{Cwiklinski2026Purification}.

The filtering, two-copy circuit constructions and the success-probability
calculation are given in Appendix~\ref{app:fixed-parity}.
The FLO implementation of the Gaussian filters is given in Appendix~\ref{app:spinor}.

\section{Full-rank obstruction and the four-mode fidelity bound}
\label{sec:fullrank}

Here, we consider a full-rank physical source on any finite number of fermionic modes.
We prove that for every such source and any fixed pure non-Gaussian target, the target fidelity of the FLO output has a uniform upper bound strictly below $1$, regardless of the number of inputs or how small a positive success probability is allowed.

We represent these sources using the Majorana attenuation channel, which allows us to analyze different noise models in a common way.
It attenuates correlations according to the number of Majorana operators they contain.
On $n$ modes, let $\gamma_S$ denote a product of $|S|$ distinct Majorana operators, and define
\begin{equation}
 \cN_{t,n}(\gamma_S)=t^{|S|}\gamma_S,\quad 0\le t\le1.
 \label{eq:main-noise-channel}
\end{equation}
These products form an operator basis, so this definition determines the action of the channel on any state.

Specifically, at $t=1$, the channel leaves the input unchanged.
As $t$ decreases, nontrivial Majorana correlations are further attenuated, while the identity term remains unchanged.
At $t=0$, the output is the maximally mixed state.
On physical states, this channel is equivalent to applying each single-Majorana error $\gamma_j$ independently with probability $(1-t)/2$.

Because the smallest eigenvalue of $\rho$ is strictly positive, $\sigma=\cN_{t,n}^{-1}(\rho)$ remains a physical density operator when $t$ is sufficiently close to $1$.
Choose some $0<t<1$ for which $\rho=\cN_{t,n}(\sigma)$ (Appendix~\ref{app:fullrank}). Thus, every full-rank physical source $\rho$ can be represented as the output of this channel.
We quantify this representation by its smallest feasible attenuation parameter.

\begin{definition}[Minimal attenuation parameter]
\label{def:attenuation-depth}
For a physical state $\rho$ on $\mathcal F_n$, define
\begin{equation}
 \begin{gathered}
 t_*(\rho)=\min\{s\in[0,1]\mid
 \rho=\cN_{s,n}(\sigma)\\
 \text{for some physical state }\sigma\}.
 \end{gathered}
 \label{eq:attenuation-depth}
\end{equation}
\end{definition}

For a normalized pure target $\ket\phi$ on $m$ modes, 
define $F_\phi(\rho_{\rm out})=\langle\phi|\rho_{\rm out}|\phi\rangle$ and write
\begin{equation}
 \begin{aligned}
 \eta_m(t)&=\left(\frac{1-t}{1+t}\right)^{2m},\\
 f_G(\phi)&=\max_{\ket g\ {\rm Gaussian}}
             |\braket{\phi|g}|^2,
 \end{aligned}
 \label{eq:main-eta}
\end{equation}
where the maximum is over normalized pure Gaussian states on $m$ modes.
In particular, $f_G(M)=1/2$.

\begin{theorem}[Full-rank obstruction and the four-mode fidelity bound]
\label{thm:fullrank}
Let $\rho\succ0$ be a physical state on $\mathcal F_n$, and set $t=t_*(\rho)<1$.

\textup{(a)} For any fixed pure non-Gaussian target $\ket\phi$ on $m$ modes,
every normalized accepted FLO output $\rho_{\rm out}$ from any finite number of
copies of $\rho$, with positive success probability, satisfies
\begin{equation}
F_\phi(\rho_{\rm out}) \le 1-\eta_m(t)\,[1-f_G(\phi)]<1.
\label{eq:main-fullrank}
\end{equation}

\textup{(b)} For the four-mode target $\ket M$, we have
\begin{equation}
 \begin{aligned}
 &\sup_{N\ge1}f_N(\rho)
 \le h(t)=\max\{1/2,a(t)\}<1,\\
 &a(t)=\Tr[P\cN_{t,4}(P)]
      =\frac{1+14t^4+t^8}{16}.
 \end{aligned}
 \label{eq:main-four-mode}
\end{equation}
\end{theorem}

We apply the four-mode bound to the white-noise source of
Sec.~\ref{sec:setting}:
\begin{equation}
 \begin{gathered}
 \omega_F=(1-\epsilon)P+\epsilon I/16,\quad 0<\epsilon<1,\\
 F=F_M(\omega_F)=1-\frac{15\epsilon}{16}\,.
 \end{gathered}
 \label{eq:main-white-source}
\end{equation}
To show that the bound can be attained, consider the state
$\rho_\tau=\cN_{e^{-2\tau},4}(P)$ obtained by applying attenuation
directly to the target, with $\tau>0$.
This family corresponds to the symmetric-poisoning model in which the
eight single-Majorana errors occur independently at equal rates;
$\tau$ is the dimensionless time obtained by absorbing their common rate.
The derivation is given in Appendix~\ref{app:geometry}.

\begin{corollary}[White-noise error floor and exact poisoning optimum]
\label{cor:noise-models}
\textup{(a)} For $0<\epsilon<8/11$, every normalized accepted FLO output
$\rho_{\rm out}$ from any finite number of copies of $\omega_F$ satisfies
\begin{equation}
 \begin{aligned}
 1-F_M(\rho_{\rm out})
 &\ge
 \frac{14(1-\sqrt{1-\epsilon})+\epsilon}{16}\\
 &\ge\frac{\epsilon}{2}
 =\frac{8(1-F)}{15}.
 \end{aligned}
 \label{eq:main-white-ceiling}
\end{equation}
\textup{(b)} For every $\tau>0$, the symmetric-poisoning source obeys
\begin{equation}
 \sup_{N\ge1}f_N(\rho_\tau)
 =\max\{1/2,F_M(\rho_\tau)\}.
 \label{eq:main-poisoning-optimum}
\end{equation}
\end{corollary}
The proof is given in Appendix~\ref{app:fullrank}.
The noise trajectory of $\rho_\tau$ and its convex-Gaussian boundary
are derived in Appendix~\ref{app:geometry}.

The white-noise family belongs to the convex-Gaussian set when
$\epsilon\ge8/11$ \cite{Oszmaniec2014Simulation}. Corollary~\ref{cor:noise-models}(a) shows that the
sources with $0<\epsilon<8/11$ are non-convex-Gaussian bound sources with respect to $\ket M$. Whether an FLO protocol can attain the square-root error lower bound
in that corollary for this family remains unknown.

This also gives a direct comparison of the inputs with the same fidelity
in Sec.~\ref{sec:setting}.
For every fixed $1/2<F<1$, Theorem~\ref{thm:fixed-parity} guarantees
that the fixed-parity source $\rho_F$ can be distilled arbitrarily close
to $\ket M$. The output infidelity of the white-noise source
$\omega_F$ with the same initial fidelity is at least $8/15$ of its
input infidelity.

\begin{figure*}[t]
 \centering
 \includegraphics[width=\textwidth]{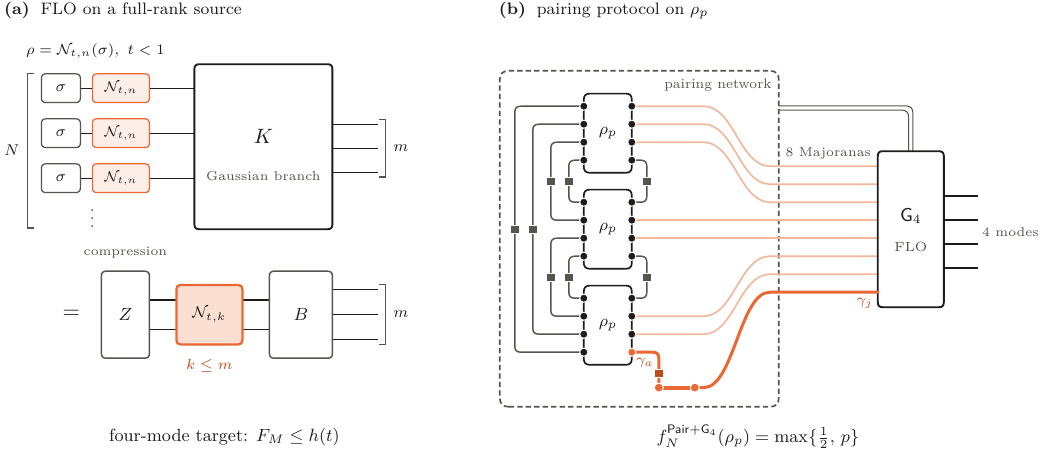}
 \caption{\textbf{Why the noise persists.}
(a) For each fully refined FLO record, compression rewrites
the output as $B\cN_{t,k}(Z)B^\dagger$
[Eq.~\eqref{eq:main-compression}],
where $Z\succeq0$ is physical and $B$ is a Gaussian map.
The same attenuation parameter $t<1$ acts on $k\le m$
effective modes, independently of the input copy number.
Each normalized branch retains a Gaussian component of weight
at least $\eta_m(t)$ (Theorem~\ref{thm:fullrank}(a)).
(b) In a $\mathsf{Pair+G}_4$ protocol
(Definition~\ref{def:pairing-protocols}), the dashed pairing
network feeds eight Majoranas into arbitrary terminal four-mode FLO.
Orange paths track single-Majorana errors to the terminal input;
the highlighted path connects $\gamma_a$ to $\gamma_j$.
Dots denote Majoranas, squares denote bilinear measurements,
and the double line denotes the classical record.
The exact optimum is given by Theorem~\ref{thm:matching}.}
\label{fig:obstructions}
\end{figure*}

Returning to a general full-rank source, we explain why increasing the
number of inputs or selecting successful records of very small
probability still cannot remove this obstruction
(Fig.~\ref{fig:obstructions}(a)).
Write $\rho=\cN_{t,n}(\sigma)$, where $t=t_*(\rho)<1$.
For $N$ input copies,
$\rho^{\otimes N}=\cN_{t,nN}(\sigma^{\otimes N})$,
so the attenuation parameter $t$ does not change with the number of inputs.

Consider an FLO protocol that leaves $m$ output modes.
After refining a successful record according to the convention of
Appendix~\ref{app:spinor}, the branch is represented by a Gaussian
Kraus operator $K$.
The compression lemma of Appendix~\ref{app:fullrank} rewrites its
unnormalized output as
\begin{equation}
 K\rho^{\otimes N}K^\dagger
 =B\cN_{t,k}(Z)B^\dagger, \quad Z\succeq0\ \text{physical},\quad k\le m.
 \label{eq:main-compression}
\end{equation}
$B$ is a Gaussian map from these effective modes to the final $m$
output modes. Both $Z$ and $B$ may depend on the inputs and the selected branch.
In this representation, attenuation retains the original parameter $t$ and acts on at most $ m$ modes, a bound set by the output size
and independent of the number of input copies.

For $t<1$, the attenuated intermediate state has a maximally mixed component with positive weight.
The maximally mixed state is itself Gaussian, so this part still contributes a Gaussian component after the Gaussian map.

Filtering changes the relative weights of the components. Comparing this weight with the total output trace of the branch shows that the normalized output of every nonzero branch contains a Gaussian component with weight at least $\eta_m(t)$.

The fidelity of a Gaussian state with the target $\ket\phi$ is at most $f_G(\phi)<1$.
Thus, even if the rest of the output were the target, this Gaussian component would leave an overall infidelity of at least $\eta_m(t)[1-f_G(\phi)]$. Every successful branch satisfies the same bound, so the bound also applies to the entire accepted output. This proves Theorem~\ref{thm:fullrank}(a).

For the four-mode target $\ket M$, four-mode spinor geometry further constrains the overlap of the entire output with the target after Gaussian filtering.
Together with the Gaussian-filter bound in Appendix~\ref{app:spinor}, this gives the tighter upper bound $h(t)$ in Theorem~\ref{thm:fullrank}(b), and hence the white-noise error estimate above.
The full proof is given in Appendix~\ref{app:fullrank}.

\section{Odd-parity noise: the exact optimum for pairing protocols}
\label{sec:matching}

This section considers the source $\rho_p=pP+(1-p)O/8$ introduced in
Sec.~\ref{sec:setting}, with $0<p<1$.
It has weight along the target and throughout the odd sector, but vanishes
on the even subspace $Q$ orthogonal to the target.
Hence, $\rho_p$ occupies both parity sectors, while having rank 9 and is
not covered by either of the preceding two theorems.

This source can be prepared from the target as follows. With probability
$p$, apply no error. With probability $1-p$, choose one of the eight
Majorana operators uniformly and apply the corresponding single-Majorana
error. The resulting state is $\rho_p$.
Unlike the attenuation in the preceding section, which allows multiple
Majorana errors to occur independently, the noise here is at most one.
For $p>0$, if an ideal nondestructive total-parity check was additionally allowed,
the even outcome would select the target from $\rho_p$, with
probability $p$, since $E\rho_pE=pP$.
This check preserves coherence within each parity sector and lies outside
our FLO operation class; Appendix~\ref{app:geometry} explains its
difference from mode-by-mode occupation measurements and gives the argument.
We ask whether processing multiple copies of $\rho_p$ can approach the
target within the original operation class. We first study this question
in a restricted FLO class that includes braid-and-fusion protocols.
Here, a pairing state is a pure joint eigenstate of the bilinears
$i\gamma_a\gamma_b$ associated with a partition of all Majoranas into
disjoint pairs. Braiding acts by signed Majorana permutations
\cite{Ivanov2001NonAbelian}.
We use these operations to define the following two-stage protocol.

\begin{definition}[Pairing protocols]
\label{def:pairing-protocols}
A protocol in $\mathsf{Pair+G}_4$ has two stages.

\textup{(i)} A \emph{pairing network} jointly processes the source copies
using independently prepared pairing ancillas, signed Majorana
permutations of determinant $+1$ on the full system and ancillary Majorana space,
and projective measurements of bilinears $i\gamma_a\gamma_b$.
Classical randomness, feed-forward, postselection, and discarding are
allowed. Four modes remain at the end of this stage.

\textup{(ii)} An arbitrary FLO operation acts on the four retained resource
modes and leaves four output modes. It may use independent Gaussian
ancillas, with the operation chosen according to the first-stage record,
but it receives no further noisy source copies.
\end{definition}

Write $f_N^{\mathsf{Pair+G}_4}$ for
Eq.~\eqref{eq:distillation-definition} with the supremum restricted to this
class. General Gaussian mixing, such as the beam splitters used in
Sec.~\ref{sec:protected}, and non-pairing Gaussian ancillas are allowed
only in the second stage.
The precise conventions are given in Appendix~\ref{app:matching}.

\begin{proposition}[Bravyi protocol containment]
\label{prop:bravyi-containment}
Every finite purification circuit in Sec.~IV of
Ref.~\cite{Bravyi2006Universal} belongs to $\mathsf{Pair+G}_4$.
With ideal operations and independent copies of $\rho_F$, the successful
outputs of its iteration approach $P$ whenever $F$ exceeds Bravyi's threshold (approximately $0.6156$).
\end{proposition}
The proof is given in Appendix~\ref{app:matching}.

Bravyi's purification analysis assumes even-parity input support
[15]. Above Bravyi's threshold, his protocol distills $\rho_F$.
At the same input fidelity, the optimal fidelity for $\rho_{p=F}$
within $\mathsf{Pair}+G_4$ remains $F$.
\begin{samepage}
\begin{theorem}[Exact pairing optimum]
\label{thm:matching}
For every $N\ge1$ and $0\le p\le1$, we have
\begin{equation}
 f_N^{\mathsf{Pair+G}_4}(\rho_p)=\max\{1/2,p\}.
 \label{eq:main-matching}
\end{equation}
Every accepted output $Y$ obtained from any finite number of copies of
$\rho_p$ by a $\mathsf{Pair+G}_4$ protocol satisfies
\begin{equation}
 \Tr[PY]\le\max\{1/2,p\}\Tr Y.
 \label{eq:main-matching-homogeneous}
\end{equation}
\end{theorem}
\end{samepage}

This value can be attained in two ways.
When $p\ge1/2$, retaining one input and discarding the remaining copies
gives fidelity $p$.
When $p<1/2$, discarding the inputs and preparing the vacuum gives
fidelity $1/2$. Processing more copies, even with the final allowed
four-mode FLO operation, cannot exceed the larger of these two values.

Consider $p\ge1/2$. Following the refinement convention of Appendix~\ref{app:spinor}, fix the classical choices and measurement outcomes in the first stage, and
include the occupation outcomes of discarded modes in the record.
Let $X$ be the nonzero unnormalized four-mode output of this record before the final FLO operation.
If $X/\Tr X$ is already convex-Gaussian, the final FLO operation preserves
this property, so its fidelity is at most $1/2$.
We therefore only need to analyze the remaining outputs.

For a fixed record, connecting the paired endpoints in the preparations,
measurements, and permutations gives a set of Majorana lines
(Fig.~\ref{fig:obstructions}(b)).
Each line starting at an output either connects to another output
Majorana or to a Majorana in a source copy.
The former case fixes the occupation of one mode.
Every physical state of the remaining three modes is convex-Gaussian,
so the normalized full output is also convex-Gaussian.
This three-mode fact was established in Appendix~\ref{app:fullrank}.
For the non-convex-Gaussian outputs that remain to be analyzed, each
output Majorana $\gamma_j$ is therefore connected by a line to a
Majorana $\gamma_a$ in some source copy.

Write $q=(1-p)/8$ and $P_a=\gamma_aP\gamma_a$.
For this source copy, use the decomposition
\begin{equation}
 \rho_p=(pP+qP_a)+q(O-P_a).
 \label{eq:source-split}
\end{equation}
Along this line, the output obtained by replacing the selected source
with $P_a$ equals the output obtained by replacing it with $P$,
conjugated by $\gamma_j$.
The two outputs have the same trace. Thus, the same record cannot retain
the target contribution in the first term while completely removing its
accompanying single-error contribution.
The coefficients $p$ and $q$ preserve the original weights of these two
contributions.

The remainder $q(O-P_a)$ can be written as a positive combination of
pairing states. In each term, the selected noisy source is replaced by an allowed pairing preparation, reducing the number of original noisy sources by one.
Induction on these terms extends the propagation relation to the full
output $X$. This decomposition can be made separately for each output Majorana to the same $X$,
giving eight constraints. 

Appendix~\ref{app:matching} proves that, for the non-convex-Gaussian
outputs under consideration, both parity blocks of $X$ are real matrices
in the spinor coordinates chosen in Appendix~\ref{app:spinor}.
In this case, combining the constraints in all eight Majorana directions
and using $8q=1-p$ gives
\[
 \lambda_{\max}(X)\le p\Tr X,
\]
where $\lambda_{\max}(X)$ is the largest eigenvalue of $X$.
For an input with real parity blocks,
Eq.~\eqref{eq:spinor-terminal} bounds the fidelity after arbitrary
subsequent Gaussian processing by
$\max\{1/2,\lambda_{\max}(X)/\Tr X\}$.
Hence, the final four-mode FLO operation cannot increase the fidelity
beyond $p$.

For $0\le p<1/2$, use
\[
 \rho_p
 =2p\,\rho_{p=1/2}+(1-2p)\frac O8.
\]
Here $O/8$ is a mixture of odd occupation states, all of which are pairing
preparations.
Expanding any finite tensor power of this decomposition, each term
contains only copies of $\rho_{p=1/2}$ and allowed pairing preparations.
The bound already proved for $p=1/2$ therefore applies to every term
that contains noisy sources.
Terms containing no noisy sources are themselves free inputs and
likewise cannot exceed fidelity $1/2$.

Apply these estimates to unnormalized outputs. Different pairing records may select different final FLO operations, but they satisfy the same homogeneous inequality, which depends only
on $p$. Summing the outputs over the accepted records gives
Eq.~\eqref{eq:main-matching-homogeneous}. Together with the two attainable choices above, this proves Theorem~\ref{thm:matching}.
The complete arguments for path propagation, positive replacement, and the structure of the four-mode output are given in Appendix~\ref{app:matching}.

The distillability of $\rho_p$ under unrestricted FLO remains unresolved.
We propose the following conjectures concerning nondistillability and the
exact optimum, respectively.

\begin{conjecture}[Odd-parity noise under FLO]
\label{conj:rank-nine}
For $0<p<1$, we conjecture the following.

\textup{(a)} $\rho_p$ is bound with respect to $\ket M$, namely,
\begin{equation}
 \sup_{N\ge1}f_N(\rho_p)<1.
 \label{eq:rank-nine-conjecture}
\end{equation}
\textup{(b)} Its optimal fidelity equals the pairing optimum,
\begin{equation}
 \sup_{N\ge1}f_N(\rho_p)=\max\{1/2,p\}.
 \label{eq:rank-nine-no-gain-conjecture}
\end{equation}
\end{conjecture}

We note that part (b) of the conjecture implies part (a).
Theorem~\ref{thm:matching} proves the equality in (b) for
$\mathsf{Pair+G}_4$ protocols.
Under general Gaussian mixing, however, a Majorana operator becomes a
linear combination of multiple Majorana operators.
The original induction, which propagates a single-Majorana error along
one line and replaces the source connected to it, therefore does not
apply directly to unrestricted FLO.
Theorem~\ref{thm:fullrank} cannot solve this question by a full-rank
approximation.
As full-rank states approach $\rho_p$, their minimum attenuation parameter
$t_*$ must tend to $1$, and the corresponding error floor tends to $0$.
Therefore, this limit does not leave a positive lower bound on the error for
$\rho_p$. A different argument is needed to treat the rank-$9$ source
under general FLO.
\section{Discussion}
\label{sec:discussion}

We have proved that, for four-mode fixed-parity sources,
non-convex-Gaussianity  characterizes distillability under fermionic linear optics (FLO).
For a physical source on any finite number of modes with full rank on the entire Fock space, we establish a positive lower bound on
the error of successful FLO outputs relative to any fixed pure
non-Gaussian target. This bound cannot be removed by increasing the
number of input copies.
Therefore, a source may retain fermionic magic without being distillable
arbitrarily close to the target.
For a given noise model and target accuracy, these bounds can help
determine whether the required accuracy exceeds the capabilities of FLO.

For the symmetric-poisoning family $\rho_\tau$ in
Corollary~\ref{cor:noise-models}(b), there is an interval in which the
source remains non-convex-Gaussian. However, even under joint processing of arbitrarily many inputs, the optimal fidelity with the GHZ state $\ket M$ is only $1/2$, the value achieved by simply preparing the vacuum.
The derivation of this noise trajectory and the corresponding interval is given in Appendix~\ref{app:geometry}.

In the example in Appendix~\ref{app:geometry}, the four-mode marginal obtained by ignoring an occupation flag is a full-rank
non-convex-Gaussian state. If the source is supplied together with this correlated flag, measuring the flag by an occupation measurement allowed by FLO and retaining a specified outcome selects a fixed-parity source $\rho_F$ with positive probability, where $F>1/2$.
Theorem~\ref{thm:fixed-parity} then applies to distill this source to arbitrary accuracy.
Thus, the unheralded marginal alone does not determine what can be accomplished with an input carrying the correlated flag.
This uses the correlated information already supplied by the source preparation.

A complete classification of distillability under general FLO remains
unavailable for sources that occupy both parity sectors but are still
rank deficient.
Whether the pairing optimum for $\rho_p $ can be exceeded when general Gaussian mixing is
allowed in the joint processing of multiple inputs is the question left open
in Conjecture~\ref{conj:rank-nine}.
For the fixed-parity sources proved to be distillable, the next step is
to find protocols that use fewer inputs at a given accuracy and to study
how imperfect operations affect the success probability and the accuracy
ultimately achievable.

\begin{acknowledgments}
The authors thank Raghav Jha and Songbo Xie for helpful discussions.
This work was supported by the U.S. Department of Energy,
Advanced Scientific Computing Research, under Contract No. DE-SC0025384. The authors acknowledge assistance from OpenAI's GPT-6 Astra and GPT-6.2 Astra with translation of an earlier Chinese draft, checking mathematical derivations, and the proof of Theorem 3, and from Anthropic's Claude Opus 5.5, Fable 5.2 with the proof of Theorem 2(b) and figure preparation. The authors have checked all results and assume full responsibility.
\end{acknowledgments}

\section*{Data Availability}
This is a theoretical study and no data were created or analyzed in this study. All results and supporting derivations are provided in the main text and the Appendices.

\makeatletter
\@booleantrue\raggedcolumn@sw
\makeatother
\onecolumngrid
\clearpage
\makeatletter
\@booleanfalse\raggedcolumn@sw
\makeatother
\normalsize
\setlength{\dbltextfloatsep}{20pt plus 2pt minus 4pt}
\setlength{\textfloatsep}{20pt plus 2pt minus 4pt}
\makeatletter
\let\set@footnotewidth\set@footnotewidth@one
\onecolumn@grid@setup
\count\footins=1000\relax
\makeatother

\appendix
\hypersetup{linkcolor=blue,citecolor=blue,urlcolor=blue}
\section{Four-mode Gaussian geometry}
\label{app:spinor}

In this section, we introduce spinor representations of the two parity sectors and their pure-state Gaussian criterion, then obtain the mixed-state concurrence and the Gaussian-filter bounds used in the subsequent proofs.

\subsection{Pure-state geometry}

Pure fermionic states on four modes have a useful geometric structure.
Their parity sectors $\mathcal H_+$ and $\mathcal H_-$ carry the two
chiral spin representations $8_s$ and $8_c$ of
$\operatorname{Spin}(8)$, while the real span of the Majoranas carries
the vector representation $8_v$. This structure gives a criterion for
whether a pure state is Gaussian. Specifically, each chiral spinor
representation has a $\operatorname{Spin}(8)$-invariant symmetric
bilinear form. For four modes, the pure Gaussian states form its null
cone. In an orthonormal basis adapted to the invariant real structure,
this form can be written as
$q(z)=z^Tz, z\in\mathbb C^8$.
A nonzero pure state $z$ is therefore Gaussian if and only if $q(z)=0$
\cite{Oszmaniec2014Simulation}. For the target state $M$, we choose
$M=e_1$ in the even sector. Then $q(M)=1$, so it is not Gaussian.

For an explicit occupation-basis convention, label the modes
$0,1,2,3$ and use
$\ket S=a_{j_1}^\dagger\cdots a_{j_k}^\dagger\ket{0000}$ for
$S=\{j_1<\cdots<j_k\}$, with $\ket\psi=\sum_S\psi_S\ket S$.
The even- and odd-sector forms are
\[
 \begin{aligned}
 q_+(\psi)=2(\psi_\varnothing\psi_{0123}
 -\psi_{01}\psi_{23}+\psi_{02}\psi_{13}-\psi_{03}\psi_{12}),\quad
 q_-(\psi)=2(\psi_0\psi_{123}
 -\psi_1\psi_{023}+\psi_2\psi_{013}-\psi_3\psi_{012}).
 \end{aligned}
\]
Writing $S^c$ for the complement of $S$ in $\{0,1,2,3\}$, the invariant
conjugation can be chosen as
\[
 \Theta\!\left(\sum_S\psi_S\ket S\right)
 =\sum_S(-1)^{\sum_{j\in S}j}\overline{\psi_S}\ket{S^c},
 \qquad q_\pm(\psi)=\langle\Theta\psi|\psi\rangle
 \quad(\psi\in\mathcal H_\pm).
\]
In the adapted spinor coordinates, $\Theta$ is the ordinary complex conjugation.

\subsection{Mixed-state criterion and inequalities}

For a mixed state, we need to determine whether it has a decomposition
consisting entirely of pure Gaussian states.
Fix a parity sector $\mathcal H_s$, with $s\in\{+,-\}$, and let
$X\succeq0$ be a positive operator in this sector.
We allow $X$ to be unnormalized so that we can directly treat
postselected outputs below.
By the pure-state criterion of the preceding subsection, we need to
determine whether there is a decomposition
$X=\sum_\alpha z_\alpha z_\alpha^\dagger$
in which every component satisfies $z_\alpha^Tz_\alpha=0$.

Fortunately, Oszmaniec, Gutt, and Ku\'s obtained a complete criterion
for this problem by applying Uhlmann's concurrence construction to
four-mode convex-Gaussianity
\cite{Uhlmann2000Fidelity,Oszmaniec2014Simulation}.
Here, we use its homogeneous form for unnormalized $X$.
In the spinor coordinates chosen above, let $V=\sqrt X$, and denote the singular
values of $V^TV$ in decreasing order by
$d_1\ge\cdots\ge d_8\ge0$.
Define
\begin{equation}
 A(X)=\sum_jd_j,\qquad d_{\max}(X)=d_1,\qquad
 C(X)=[2d_{\max}(X)-A(X)]_+,
 \label{eq:spinor-invariants}
\end{equation}
where $[x]_+=\max(x,0)$.
For any ensemble factorization $X=ZZ^\dagger$, the matrix $Z^TZ$
gives the same nonzero singular values.

We write $\operatorname{cone}(\CG_4)$ for the set consisting of the zero operator and all
positive operators whose normalized states belong to $\CG_4$.
For membership in this cone, a single-sector operator is extended by
zero on the opposite sector.
The concurrence formula relates these singular values to the
decomposition problem:
\begin{equation}
 C(X)=\inf_{X=\sum_\alpha z_\alpha z_\alpha^\dagger}
       \sum_\alpha|z_\alpha^Tz_\alpha|,
 \qquad
 X\in\operatorname{cone}(\CG_4)\ \Longleftrightarrow\ C(X)=0.
 \label{eq:spinor-concurrence}
\end{equation}
Thus, the existence of a decomposition entirely into Gaussian vectors
reduces to comparing $d_1$ with the sum of the other seven singular values.

For real parity blocks, the calculation simplifies further.
If $X$ is real in these spinor coordinates, then $V^TV=X$, so
\[
 A(X)=\Tr X,\qquad
 d_{\max}(X)=\lambda_{\max}(X).
\]
We can use these quantities to estimate the fidelity after filtering.

\begin{lemma}[Basic properties of $A$]
\label{lem:spinor-assistance-order}
For a positive operator $X$ in these spinor coordinates, we have
\begin{equation}
 A(X)=\|\overline V V\|_1
     =\Tr\sqrt{V\overline X V}\le\Tr X,
 \qquad V=\sqrt X.
 \label{eq:spinor-root-fidelity}
\end{equation}
Furthermore, $A(cX)=cA(X)$ for $c\ge0$.
\end{lemma}

\begin{proof}
Since $A(X)$ is the sum of the singular values of $V^TV$,
$A(X)=\|V^TV\|_1$.
The matrix $V$ is Hermitian, so $V^T=\overline V$, and the definition
of the trace norm gives
\[
 A(X)=\|\overline V V\|_1
     =\Tr\sqrt{V\overline X V}.
\]
Applying the Schatten--H\"older inequality to the two factors gives
\[
 A(X)\le\|\overline V\|_2\|V\|_2=\Tr X.
\]
Replacing $X$ by $cX$ multiplies $V^TV$ and all its singular values
by $c$, hence $A(cX)=cA(X)$.
\end{proof}

Having estimated the total weight, we also need to control the output
weight in the target direction.
We have chosen $M=e_1$ in the even sector, so a pure vector $z$ has
weight $|z_1|^2$ in that direction.
The triangle inequality gives
$2|z_1|^2\le\|z\|^2+|z^Tz|$.
Summing over the vectors in any decomposition
$X=\sum_\alpha z_\alpha z_\alpha^\dagger$
and taking the infimum over all decompositions yields
\begin{equation}
 2\Tr(PX)\le\Tr X+C(X)\qquad(X\text{ even}).
 \label{eq:spinor-overlap}
\end{equation}
For even convex-Gaussian $X$, $C(X)=0$.
Thus, $W=E-2P$ satisfies $\Tr(WX)\ge0$.
This is the witness used below to treat filtering of singular inputs.

\subsection{Gaussian filters and output bounds}

A map $L$ has a definite parity if
$\Pi_{\rm out}L=\pm L\Pi_{\rm in}$, where $\Pi_{\rm in}$ and
$\Pi_{\rm out}$ are the input and output parity operators.
The positive sign preserves the parity sectors, while the negative
sign exchanges them.
Gaussian unitaries act as $\mathrm{SO}(8)$
in each spinor sector, and invertible Gaussian filters give the
complexified projective action $\mathrm{SO}(8,\mathbb C)$.
Every such sector transformation has a Gaussian spin lift on the full
Fock space, which becomes a Kraus contraction after suitable rescaling.
By polar decomposition and diagonalization of the
Hermitian quadratic factor, every invertible parity-preserving Gaussian
filter $K$ can be written as $K=U_1DU_2$, where $U_1,U_2$ are Gaussian
unitaries and $D$ is the nonunitary diagonal part,
\begin{equation}
 D=c_0\bigotimes_{j=1}^4F_j,\quad
 F_j=\operatorname{diag}(1,\nu_j)\ \text{or}\ 
     \operatorname{diag}(\nu_j,1),\quad 0<\nu_j\le1.
 \label{eq:spinor-attenuation}
\end{equation}

Therefore, it suffices to implement the middle factor $D$ to realize $K$.
Each $F_j$ leaves one of the $\ket0$ and $\ket1$ components unchanged
and multiplies the other by $\nu_j$.
Different occupation configurations select different diagonal entries
of $F_j$ on each mode and consequently acquire different overall multipliers.
For example, for $F_j=\operatorname{diag}(1,\nu_j)$, mix the
system mode with a vacuum ancilla on a beam splitter of amplitude
transmittance $\nu_j$.
If both modes are empty, the beam splitter only mixes the two modes
and cannot create particles, so measuring the ancilla gives the empty
outcome with certainty. If the system is occupied, this operation
produces a superposition. Selecting a term with an empty ancilla
implements the desired $F_j$.
Using an occupied ancilla and selecting the occupied outcome instead
gives $\operatorname{diag}(\nu_j,1)$ \cite{Bravyi2005Lagrangian}.
Applying these operations mode by mode yields $K$ with a suitable
overall rescaling. Parity-changing filters also require a Majorana
operator, whose FLO implementation is given in
Appendix~\ref{subsec:parity-conversion}.

We ask how large the successful output fidelity can be for any such
filter. We address this by comparing $A$ and $C$ before and after
filtering. Both quantities are constructed from $q$, so we first
calculate how filtering changes $q$.

The operator $D$ is diagonal of products of amplitudes for complementary configurations.
On mode $j$, one member of each complementary pair is empty and the
other is occupied, so they select the two diagonal entries $1$ and
$\nu_j$ of $F_j$. Including all four modes and the overall factor
$c_0$, every paired term acquires the same multiplier
$c_0^2\prod_j\nu_j$.
This calculation holds in both the even and odd sectors, so $D$
rescales their quadratic forms by the same overall factor.
The Gaussian unitaries on either side preserve $|q|$,
as does the Majorana factor used for parity conversion.
Restoring these factors therefore leaves the common absolute scale unchanged.

For any invertible definite-parity Gaussian map $L$, let $L_s$
be its restriction to input sector $s$. Every vector $z$ in that sector satisfies
$q(L_sz)=\zeta_s q(z)$, where $\zeta_s$ is independent of $z$.
In the chosen spinor coordinates,
$q(L_sz)=z^TL_s^TL_sz$, hence
\begin{equation}
 L_s^TL_s=\zeta_s I,\qquad
 |\zeta_+|=|\zeta_-|=c>0.
 \label{eq:spinor-common-scale}
\end{equation}
If $L$ is odd, the input and output sectors of $L_s$ are opposite.
In particular, a Majorana operator has $c=1$ and therefore preserves $A$:
\begin{equation}
 A(\gamma_jX_s\gamma_j)=A(X_s),
 \qquad s\in\{+,-\}.
 \label{eq:spinor-majorana-assistance}
\end{equation}

The common scale factor $c$ appears in the estimates of both the
target overlap and the total trace, so the fidelity upper bound need
only depend on the fixed input.

\begin{lemma}[Common Gaussian-filter bound]
\label{lem:spinor-filter-bound}
Let $X=X_+\oplus X_-\succeq0$ be a four-mode physical operator, and
write $A_s=A(X_s)$.
For any definite-parity Gaussian Kraus operator $L$ from four modes
to four modes, a nonzero output $Y=LXL^\dagger$ satisfies
\begin{equation}
 \frac{\Tr(PY)}{\Tr Y}
 \le C_X:=\max\left\{\frac12,
       \frac{\max\{d_{\max}(X_+),d_{\max}(X_-)\}}
            {A_++A_-}\right\}.
 \label{eq:spinor-filter-bound}
\end{equation}
Set $C_X=1/2$ if $A_++A_-=0$.
The same bound holds for any accepted Gaussian operation on $X$
that retains four output modes and has nonzero output.
\end{lemma}

\begin{proof}
First suppose that $L$ is invertible and $A_++A_->0$.
Set $V_s=\sqrt{X_s}$, so the corresponding output block has the
factorization
$L_sX_sL_s^\dagger=(L_sV_s)(L_sV_s)^\dagger$.
By Eq.~\eqref{eq:spinor-common-scale},
\[
 (L_sV_s)^T(L_sV_s)
 =V_s^TL_s^TL_sV_s
 =\zeta_sV_s^TV_s.
\]
Therefore, the spinor singular values of each output block are those
of the corresponding input block multiplied by $c$.
The quantities $A$ and $C$ of both blocks also scale by this common factor.

Use $s$ for the input sector sent to the even output and $o$ for the
other sector.
Equation~\eqref{eq:spinor-root-fidelity} gives the trace lower bounds
$cA_s$ and $cA_o$ for the two blocks.
Applying Eq.~\eqref{eq:spinor-overlap} to the even output gives
$2\Tr(PY)\le\Tr Y_++cC(X_s)$.
After subtracting the total trace, the odd-output contribution is
$-\Tr Y_-\le-cA_o$, so
\begin{equation}
 \Tr Y\ge c(A_s+A_o),\qquad
 2\Tr(PY)-\Tr Y
 \le c\bigl(C(X_s)-A_o\bigr).
 \label{eq:spinor-excess}
\end{equation}
The first inequality controls the normalization denominator, and the
second controls the target weight in excess of half the total weight.

By $C(X_s)=[2d_{\max}(X_s)-A_s]_+$ and the definition of $C_X$, we have
\[
 C(X_s)-A_o\le(2C_X-1)(A_s+A_o),
 \qquad 2C_X-1\ge0.
\]
When $C(X_s)=0$, the left-hand side is nonpositive.
If $C(X_s)>0$, the left-hand side is
$2d_{\max}(X_s)-(A_s+A_o)$, and
$d_{\max}(X_s)\le C_X(A_s+A_o)$.
Substituting this estimate into Eq.~\eqref{eq:spinor-excess} and
using the lower bound on the total trace gives
\[
 2\Tr(PY)-\Tr Y
 \le c(2C_X-1)(A_s+A_o)
 \le(2C_X-1)\Tr Y.
\]
Rearranging yields $\Tr(PY)\le C_X\Tr Y$.

When $A_++A_-=0$, all spinor singular values of both blocks vanish,
so both have zero concurrence.
By Eq.~\eqref{eq:spinor-concurrence}, $X$ belongs to the
convex-Gaussian cone. Gaussian processing preserves this cone, so
every nonzero output has fidelity at most $1/2$.

For singular $L$, we approximate it by invertible maps of the above form.
Let us replace every zero $\nu_j$ in its Gaussian normal form by positive
values tending to zero, while keeping the other factors fixed.
This gives invertible filters $L^{(n)}\to L$ of the same parity.
Both sides of the homogeneous inequality
$\Tr(PLXL^\dagger)\le C_X\Tr(LXL^\dagger)$
are continuous in $L$, so it remains valid in the limit.
This form also includes zero-output branches.

A general accepted operation may choose different operations according
to earlier outcomes. Once the classical choices and the entire record
are fixed, the operations along each branch are determined.
If an ancilla is a mixed Gaussian state, we decompose it into a mixture
of pure Gaussian states so that each refined branch has a specified
pure Gaussian ancillary preparation.
Next, we resolve the discarded modes into occupation outcomes. Then each
refined branch corresponds to a definite-parity Gaussian Kraus
operator, and the original accepted output is a positive sum or
integral of these branch outputs.
All branches act on the same input $X$ and therefore have the common
upper bound $C_X$. Summing or integrating their homogeneous inequalities
gives the bound for the accepted output.
\end{proof}

For nonzero $X$ with both parity blocks real, $V_s^TV_s=X_s$,
so the spinor singular values are the eigenvalues of the blocks.
In this case,
\[
 A_++A_-=\Tr X,\qquad
 \max\{d_{\max}(X_+),d_{\max}(X_-)\}
 =\lambda_{\max}(X).
\]
The upper bound in the lemma reduces to
$\max\{1/2,\lambda_{\max}(X)/\Tr X\}$.

We choose a normalized real eigenvector of definite parity corresponding
to the largest eigenvalue.
If it lies in the odd sector, we first apply the parity conversion in
Appendix~\ref{subsec:parity-conversion}, then use the
$\mathrm{SO}(8)$ action of Gaussian unitaries to map it to $M$.
Applying this transformation to the whole operator $X$ preserves its
total trace and makes the weight in the target direction
$\lambda_{\max}(X)$, giving fidelity
$\lambda_{\max}(X)/\Tr X$.
Another allowed choice is to discard the input and prepare the vacuum,
whose target fidelity is $1/2$.
Choosing the operation with the higher fidelity attains the bound; hence
\begin{equation}
 F_G^*(X)=\max\left\{\frac12,
                       \frac{\lambda_{\max}(X)}{\Tr X}\right\},
 \label{eq:spinor-terminal}
\end{equation}
where $F_G^*(X)$ denotes the supremum fidelity obtainable by arbitrary
Gaussian processing of a single input $X/\Tr X$, including discarding
the input and preparing a Gaussian output.
\section{Distillation in a fixed parity sector}
\label{app:fixed-parity}

To prove the distillable direction of
Theorem~\ref{thm:fixed-parity}, we use Gaussian
filtering to obtain an output with fidelity greater than $1/2$, then
apply a finite Gaussian twirl that preserves the target fidelity and
puts it in the standard family $\rho_F$. Finally, a two-copy circuit
repeatedly increases its target-to-noise ratio.

\subsection{Exposing the resource}
\label{subsec:parity-conversion}

First, suppose that $\rho$ has full rank in the even sector and
$C(\rho)>0$, so this even input is non-convex-Gaussian.
By the definition of $C(\rho)$, the concurrence condition gives
$d_1>\sum_{j>1}d_j$.
In the chosen spinor coordinates, $M=e_1$, so the first diagonal entry
is the target weight. Thus, filtering $\rho$ into an output with
diagonal weights $d_1,\ldots,d_8$ gives target fidelity greater than $1/2$.

Set $V=\rho^{1/2}$ and take the Takagi factorization to construct the
required filter:
\[
 V^TV=U\Delta U^T,\qquad
 \Delta=\operatorname{diag}(d_1,\ldots,d_8).
\]
Full rank of $\rho$ makes $V^TV$ invertible; hence, $\Delta\succ0$.
Since $(V\overline U)^T(V\overline U)=\Delta$, normalizing its
columns with respect to this bilinear form gives
\[
 H=V\overline U\Delta^{-1/2},\qquad H^TH=I.
\]
The choice must ensure that $H\in\mathrm{SO}(8,\mathbb C)$.
If $\det H=-1$, we replace $U$ by $US$ with
$S=\operatorname{diag}(-1,1,\ldots,1)$.
Since $S\Delta S=\Delta$, the Takagi factorization is unchanged,
while $H$ becomes $HS$ and has a determinant $+1$.
Then, by Appendix~\ref{app:spinor}, its inverse $g=H^{-1}$
corresponds to a Gaussian filter and satisfies
\begin{equation}
 g\rho g^\dagger=\Delta,\qquad
 \frac{\Tr(Pg\rho g^\dagger)}{\Tr(g\rho g^\dagger)}
 =\frac{d_1}{\sum_jd_j}>\frac12.
 \label{eq:spinor-exposure}
\end{equation}
Multiplying its Gaussian lift by a sufficiently small nonzero scalar
gives a valid Kraus operator.

For singular even $\rho$, still assume $C(\rho)>0$.
For sufficiently small $\delta>0$, $\rho+\delta E$ has full rank
in the even sector and, by continuity of the concurrence,
$C(\rho+\delta E)>0$.
Applying the preceding construction to this positive operator gives
a Gaussian filter $g$ such that
\[
 \Tr\!\left[Wg(\rho+\delta E)g^\dagger\right]<0,
 \qquad W=E-2P.
\]

Let us apply this filter to the original $\rho$.
Since $E/8$ is convex-Gaussian and Gaussian filtering preserves the cone,
$gEg^\dagger$ belongs to the cone. The witness property of $W$ gives
$\Tr(WgEg^\dagger)\ge0$, and therefore
\begin{equation}
 \begin{aligned}
 \Tr(Wg\rho g^\dagger)
 =\Tr\!\left[Wg(\rho+\delta E)g^\dagger\right] -\delta\Tr(WgEg^\dagger)<0.
 \end{aligned}
 \label{eq:spinor-singular-exposure}
\end{equation}
Thus, the same $g$, acting on the original singular input, also gives
a nonzero output with target fidelity greater than $1/2$.

If the input lies in the odd sector, we first convert it reversibly to
the even sector. Add a vacuum ancillary mode, choose a system Majorana
$\gamma$ and an ancillary Majorana $\xi$, and apply
\[
 U=\exp[(\pi/2)\gamma\xi]=\gamma\xi.
\]
For a fixed-parity input, the ancillary mode becomes occupied
deterministically, while the system action is $\gamma$ up to an
irrelevant phase. This Gaussian unitary therefore converts the odd
input reversibly to the even sector.

Finally, we apply the finite Gaussian twirl of
Appendix~\ref{app:geometry} to the resulting state with $F>1/2$.
This twirl preserves the target fidelity and puts the state in the
standard form $\rho_F$ of Eq.~\eqref{eq:even-family}.

\subsection{The circuit and its recurrence}

We construct an explicit two-copy protocol that further increases
the target-to-noise ratio and, upon iteration, brings the fidelity
arbitrarily close to $1$.

Let us label the two sets of input modes by $0,\ldots,3$ and $4,\ldots,7$.
First, we apply a passive Gaussian unitary and retain the four output modes
\begin{equation}
 b_0 =(a_0-a_5)/\sqrt2,\quad b_1 =(a_1+a_4)/\sqrt2,\quad b_2 =(a_2-a_7)/\sqrt2, \quad b_3 =(a_3+a_6)/\sqrt2,
 \label{eq:rec-modes}
\end{equation}
postselecting the remaining modes onto vacuum. Then we apply the Gaussian
filter
\begin{equation}
 L=2^{(\widehat N_{\mathrm{out}}-4)/4}
   =\bigotimes_{j=0}^3\operatorname{diag}(2^{-1/4},1)
 \label{eq:rec-filter}
\end{equation}
on the four retained modes. By Eq.~\eqref{eq:spinor-attenuation}, each
single-mode factor is implemented with an occupied Gaussian ancilla, a
beam splitter of amplitude transmittance $2^{-1/4}$, and occupied
ancillary postselection. Hence, the whole circuit belongs to the
allowed FLO operations.

This choice of filter maps the ideal input back to the target itself.
For $\ket M^{\otimes2}$, before applying $L$, the accepted state's
vacuum amplitude is twice its fully occupied amplitude. The filter
halves the former and leaves the latter unchanged, so the final
accepted output is proportional to $\ket M$.

We calculate the action of this protocol on general
$X=\alpha P+\beta Q$. Define the real matrix $R$ by $b=Ra/\sqrt2$ in
Eq.~\eqref{eq:rec-modes}. In increasing occupation order, the
pre-twirl Kraus matrix is
\begin{equation}
 K_{S,I}=
 \begin{cases}
 \frac12\,2^{-|S|/4}\det R_{S,I},&|S|=|I|,\\
 0,&|S|\ne|I|,
 \end{cases}
 \label{eq:rec-kraus}
\end{equation}
where $S\subseteq\{0,1,2,3\}$, $I\subseteq\{0,\ldots,7\}$,
and the empty determinant is one.
In particular,
\[
 K\ket M^{\otimes2}
 =\frac{\ket{0000}+\ket{1111}}{4}
 =\frac{\ket M}{2\sqrt2}.
\]
These minors directly give the following three
input contributions:
\[
 \begin{array}{c|c|c}
 \text{input operator}&\text{target accepted weight}&\text{total accepted weight}\\ \hline
 P\otimes P&1/8&1/8\\
 P\otimes Q+Q\otimes P&1/16&7/8\\
 Q\otimes Q&3/16&5/4
 \end{array}
\]
These contributions completely determine the action of the protocol on
any $X=\alpha P+\beta Q$. Set
\[
 Y=K(X\otimes X)K^\dagger.
\]
Then the only potentially nonzero entries $Y_{S,T}=\bra S Y\ket T$ are
\[
 \begin{aligned}
 Y_{\varnothing,\varnothing}=\frac{(\alpha+\beta)^2}{16},\quad
 Y_{0123,0123}=\frac{\alpha^2+7\beta^2}{16},\quad
 Y_{\varnothing,0123}=Y_{0123,\varnothing} =\frac{\alpha^2-\beta^2}{16},\quad
 Y_{S,S}=\frac{\beta(\alpha+\beta)}8\quad(|S|=2).
 \end{aligned}
\]

After applying the target-preserving twirl $\mathcal T$ of
Eq.~\eqref{eq:geometry-twirl} to $Y$, the output again has the form
$\widetilde\alpha P+\widetilde\beta Q$.
The twirl preserves the target overlap and total trace of $Y$ and
distributes the remaining even-sector weight uniformly over the seven
directions supported by $Q$.
The coefficients after twirling are therefore
$\widetilde\alpha=\Tr(PY)$ and
$\widetilde\beta=[\Tr Y-\Tr(PY)]/7$.
Substituting the entries of the table gives
\begin{equation}
 \mathcal T(Y)
 =\widetilde\alpha P+\widetilde\beta Q,\quad
 \widetilde\alpha
=\frac{2\alpha^2+\alpha\beta+3\beta^2}{16},
 \quad
 \widetilde\beta
 =\frac{\beta(13\alpha+17\beta)}{112}.
 \label{eq:rec-homogeneous}
\end{equation}

The input and accepted output belong to the same family, and their
fidelity is determined by the ratio of the target weight to the weight
of each noise direction.
For $\beta>0$, set
\[
 r=\frac{\alpha}{\beta},
 \quad
 F=\frac{\alpha}{\alpha+7\beta}=\frac{r}{r+7}.
\]
Notice that the fidelity $F$ is strictly increasing in $r$, and $F>1/2$ is
equivalent to $r>7$.
Taking the ratio of the two coefficients in
Eq.~\eqref{eq:rec-homogeneous} gives
\[
r'=\frac{\widetilde\alpha}{\widetilde\beta}
     =\frac{7(2r^2+r+3)}{13r+17},\quad
 r'-r=\frac{(r-7)(r-3)}{13r+17},
\]
which are Eqs.~\eqref{eq:main-recurrence}
and~\eqref{eq:main-gain} in the main text.

When the initial $r>7$, every round has $r'>r$, so the iterates increase.
If this increasing sequence were bounded, it would converge to a finite
fixed point greater than $7$.
But the only finite fixed points of this recurrence are $3$ and $7$,
so iteration must give $r\to\infty$ and hence $F\to1$.
If $\beta=0$, the normalized input is already $P$, and no further
distillation is needed.

For input states $X/\Tr X$, the success probability of one round is
\begin{equation}
 s(r)=\frac{\widetilde\alpha+7\widetilde\beta}
 {(\alpha+7\beta)^2}
 =\frac{(r+2)(r+5)}{8(r+7)^2}>0.
 \label{eq:rec-success}
\end{equation}
Since $F\to1$, any prescribed target fidelity $1-\varepsilon$
can be reached in finitely many rounds.
Each round processes two outputs from the preceding round, so a fixed
finite number of rounds requires only finitely many initial inputs.
All of these steps have positive success probabilities. Together with
the initial single-copy filtering, they give a finite protocol with
positive total success probability.

In contrast, if the initial state is convex-Gaussian, every tensor power
remains convex-Gaussian, and FLO with postselection preserves the
convex-Gaussian cone \cite{Bravyi2005Lagrangian}.
By Appendix~\ref{app:spinor}, all nonzero accepted outputs have target
fidelity at most $1/2$.
Discarding the input and preparing the vacuum attains $1/2$, so
$f_N(\rho)=1/2$ for every $N\ge1$.
Together with the parity conversion above, this completes the proof
of Theorem~\ref{thm:fixed-parity} for both parity sectors.
\section{Full-rank obstruction and the four-mode fidelity upper bound}
\label{app:fullrank}

In this section, we prove Theorem~\ref{thm:fullrank} and Corollary~\ref{cor:noise-models}.
The strategy is first to find an attenuation representation of a full-rank source that works for every copy number.
We then use Gaussian branch compression to restrict the residual attenuation that needs to be estimated to at most $m$ modes. The successful output therefore always retains a convex-Gaussian component whose weight has a lower bound independent of the copy number. Since the fidelity of this component with a pure non-Gaussian target is strictly less than one, the whole output cannot approach that target arbitrarily closely.

\subsection{A tensor-stable attenuation parameter}

Let $\rho\succ0$ be a physical state on the full $\mathcal F_n$.
We seek a $t<1$ such that $\rho$ is the output of a physical state under $\cN_{t,n}$.
For $t>0$, $\sigma_t=\cN_{t,n}^{-1}(\rho)$ determines the required preimage.
This inverse map preserves Hermiticity, trace, and parity, so it remains only to check that $\sigma_t$ is positive.

At $t=1$, the channel is the identity, so $\sigma_1=\rho$.
Here, the full-rank condition allows positivity to be retained:
the smallest eigenvalue of $\rho$ is strictly positive, and a sufficiently small perturbation cannot make it negative.
Thus, $\sigma_t$ remains a physical density operator when $t<1$ is sufficiently close to $1$.
Choose such a $0<t<1$ and write $\sigma=\sigma_t$.
This representation also applies to multiple inputs.
On disjoint mode sets, the degrees of Majorana monomials add under multiplication,
so attenuation factorizes on physical tensor products, giving
\begin{equation}
 \rho=\cN_{t,n}(\sigma),\qquad
 \rho^{\otimes N}=\cN_{t,nN}(\sigma^{\otimes N})
 \quad(N\ge1).
 \label{eq:noise-source-preimage}
\end{equation}
Hence, the same $t$ can be retained as the copy number increases.

The compression below requires moving attenuation past Gaussian unitaries, which follows because Gaussian unitaries preserve Majorana degree.
We also need to extract a quantitative lower bound from attenuation,
so we write it in random-unitary form.
Set $R_j=i\Pi_m\gamma_j$, $u=(1+t)/2$, and $v=(1-t)/2$.
Write $\operatorname{Ad}_{R_j}(X)=R_jXR_j^\dagger$ for conjugation by $R_j$.
Each $R_j$ is a Hermitian unitary whose conjugation changes the sign of $\gamma_j$ and leaves the other Majoranas unchanged.
These conjugations commute with each other.
Since $u-v=t$ and $u+v=1$,
each factor only multiplies monomials containing $\gamma_j$ by $t$, giving
\begin{equation}
 \cN_{t,m}
 =\prod_{j=1}^{2m}
 \left(u\,\mathrm{id}+v\,\operatorname{Ad}_{R_j}\right).
 \label{eq:noise-mixture}
\end{equation}
This representation holds on the full operator algebra and gives a self-adjoint random-unitary channel.
On physical operators, $R_jXR_j=\gamma_jX\gamma_j$, so it also describes independent single-Majorana errors of probability $v$.\footnote{For $t>0$, $\cN_t$ is the fermionic oscillator semigroup at time $-\log t$
\cite{Carlen1993Hypercontractivity}. On physical operators, its generator is a special case of dissipative FLO with linear Lindblad
operators \cite{Bravyi2012Dissipative}.}

To use the smallest parameter in Eq.~\eqref{eq:attenuation-depth}, we must also show that this minimum is attained.
The pairs $(s,\sigma)$ satisfying $\rho=\cN_{s,n}(\sigma)$
form a closed subset of the compact product of $[0,1]$ and the physical state space,
and this subset contains $(1,\rho)$.
Therefore, the feasible set is nonempty and compact, and the parameter $s$ attains its minimum on it.

Equation~\eqref{eq:noise-mixture} also shows why a source with a feasible parameter less than $1$ must have full rank.
For $0\le s<1$, expanding this equation on $n$ modes gives a weight of at least $[(1-s)/2]^{2n}$ for each unitary product.
These products form a complete orthogonal operator basis,
and the uniform average of their conjugations sends every state to $I/2^n$.
Extracting this common minimum weight from each term gives
\[
 \cN_{s,n}(\sigma)
 \succeq \frac{(1-s)^{2n}}{2^n}I
 \succ0.
\]
Together with the full-rank preimage construction above, this gives
\[
 t_*(\rho)<1
 \quad\Longleftrightarrow\quad
 \rho\succ0.
\]
Since $\eta_m(t)$ is nonincreasing in $t$ and $h(t)$ is nondecreasing in $t$,
the smallest feasible parameter $t_*(\rho)$ gives the strongest bounds supplied by this method.
In what follows, we take $t=t_*(\rho)$ and fix a physical preimage $\sigma$ satisfying
$\rho=\cN_{t,n}(\sigma)$.

\subsection{Compression and a uniform Gaussian residue}
The idea is first to compress the action of each branch to at most $m$ effective modes,
and then to estimate the attenuation on these modes. We first consider a definite-parity Gaussian branch under the refined-branch convention of Appendix~\ref{app:spinor}.

\begin{lemma}[Compression and a uniform Gaussian residue]
\label{lem:noise-compression}
\label{lem:noise-uniform-residue}
Let $0\le t<1$, let $X\succeq0$ be physical and let
$K:\mathcal F_{nN}\to\mathcal F_m$ be a nonzero
Gaussian branch of definite parity. Its output $Y_K:=K\cN_{t,nN}(X)K^\dagger$ can be compressed to at most $m$ effective modes.
There exist $0\le k\le\min(nN,m)$, a Gaussian map of definite parity
$B:\mathcal F_k\to\mathcal F_m$, and a physical $Z\succeq0$ such that
\begin{equation}
 Y_K=B\cN_{t,k}(Z)B^\dagger,
 \qquad BB^\dagger=KK^\dagger.
 \label{eq:noise-conditional-representation}
\end{equation}
The state $\tau_K=KK^\dagger/\Tr(KK^\dagger)$ is Gaussian and obeys the following weight bound in the output:
\begin{equation}
 Y_K\succeq\eta_m(t)\Tr(Y_K)\tau_K,
 \qquad
 \eta_m(t)=\left(\frac{1-t}{1+t}\right)^{2m}>0.
 \label{eq:noise-uniform-residue}
\end{equation}
The coefficient depends only on $t$ and the number $m$ of output modes.
\end{lemma}

\begin{proof}
First, we identify the support of $K$.
Gaussian maps are closed under adjoints and composition
\cite{Bravyi2005Lagrangian}. Applying $K$ and $K^\dagger$ to maximally mixed Gaussian inputs shows that $KK^\dagger$ and $K^\dagger K$ are positive Gaussian operators.
After diagonalizing $K^\dagger K$ into one-mode factors, each factor has rank one or two. Hence, its support projection can be written as
\begin{equation}
 E_K=U^\dagger(I_{\mathcal F_k}\otimes\ket g\bra g)U,
 \label{eq:noise-support}
\end{equation}
where $\ket g$ is an occupation state on the remaining $nN-k$
modes.
Since $\operatorname{rank}K=\operatorname{rank}(K^\dagger K)=2^k$
and $K$ maps into $\mathcal F_m$, we have $2^k\le2^m$, hence $k\le m$.

Thus, after $U$, $K$ only needs to act on $k$ effective input modes,
while the other modes are fixed in $\ket g$.
Set $C=(I\otimes\bra g)U$ to remove these fixed modes
and denote the remaining action by $B=KC^\dagger$.
Since $C^\dagger C=E_K$ and $K=KE_K$, we get
\begin{equation}
 K=BC,\qquad CC^\dagger=I,\qquad BB^\dagger=KK^\dagger.
 \label{eq:noise-compression-factor}
\end{equation}
The map $C^\dagger=U^\dagger(I\otimes\ket g)$ is a Gaussian
preparation isometry of definite parity, so $B$ remains an allowed
Gaussian branch of definite parity.

This compression retains the original attenuation parameter $t$.
Using commutation of $\cN_t$ with $U$, put $X'=UXU^\dagger$
and order the retained modes $\mathsf S$ before the remaining modes $\mathsf R$.
Write $\Pi_{\mathsf S}$ and $\Pi_{\mathsf R}$ for their respective total parity operators.
Since $\ket g\bra g$ is physical, its contraction
eliminates terms of odd environment degree, while physicality of $X'$ ensures that the surviving terms also have even system degree.
Hence, attenuation factorizes over $\mathsf S$ and $\mathsf R$ on these terms.

Self-adjointness of attenuation then moves the channel on $\mathsf R$ to the projector $\ket g\bra g$. This gives the new preimage
\begin{equation}
H=\cN_{t,nN-k}(\ket g\bra g),\quad
 Z=\Tr_{\mathsf R}[(I\otimes H)X']
  =\Tr_{\mathsf R}\!\left[
      (I\otimes\sqrt H)X'(I\otimes\sqrt H)\right].
 \label{eq:noise-conditioned-preimage}
\end{equation}
Moreover, $[X',\Pi_{\mathsf S}\otimes\Pi_{\mathsf R}]=0$ and $[H,\Pi_{\mathsf R}]=0$, together with partial-trace cyclicity,
give $[Z,\Pi_{\mathsf S}]=0$, so $Z$ is physical.
The above contraction gives
\[
 C\cN_{t,nN}(X)C^\dagger
 =
 (I\otimes\bra g)\cN_{t,nN}(X')(I\otimes\ket g)
 =
 \cN_{t,k}(Z).
\]
Together with $K=BC$, this proves
Eq.~\eqref{eq:noise-conditional-representation}.

Finally, we estimate both an operator lower bound and the total output weight
to determine the Gaussian fraction that remains after normalization.
On these $k$ effective modes, let $R_S$ denote products of the reflections.
Up to phases, they form a complete orthogonal unitary
operator basis, so
\[
 \sum_{S\subseteq[2k]}R_SZ R_S^\dagger
 =2^k\Tr(Z)I.
\]
The corresponding weights in Eq.~\eqref{eq:noise-mixture}
lie between $v^{2k}$ and $u^{2k}$, giving
\begin{equation}
 2^kv^{2k}\Tr(Z)I
 \preceq\cN_{t,k}(Z)
 \preceq2^ku^{2k}\Tr(Z)I.
 \label{eq:noise-sandwich}
\end{equation}
Conjugate both inequalities by $B$, take the trace of the upper bound,
and use $BB^\dagger=KK^\dagger$ to obtain
\[
Y_K\succeq2^kv^{2k}\Tr(Z)\,KK^\dagger,\quad \Tr Y_K\le2^ku^{2k}\Tr(Z)\,\Tr(KK^\dagger).
\]
Comparing these estimates gives
\[
 Y_K\succeq
 \left(\frac vu\right)^{2k}\Tr(Y_K)\tau_K.
\]
Since $0<v/u\le1$ and $k\le m$, this gives
Eq.~\eqref{eq:noise-uniform-residue}.
Gaussianity of $KK^\dagger$ ensures that $\tau_K$ is a Gaussian state.
These estimates also cover $k=0$, $t=0$, and zero output.
\end{proof}

This uniform lower bound on the Gaussian weight already gives a fidelity gap for a general target.
Put $P_\phi=\ket\phi\bra\phi$, where $\ket\phi$ is a fixed
pure non-Gaussian target.
Pure Gaussian states on fixed $m$ modes form a compact set,
so $\Tr(P_\phi\tau_K)\le f_G(\phi)<1$.
Taking the trace against $I-P_\phi$ in Eq.~\eqref{eq:noise-uniform-residue} gives
\begin{equation}
 \Tr[(I-P_\phi)Y_K]
 \ge\eta_m(t)[1-f_G(\phi)]\Tr Y_K.
 \label{eq:noise-target-bound}
\end{equation}
Consider $X=\sigma^{\otimes N}$ from
Eq.~\eqref{eq:noise-source-preimage}. Every refined branch satisfies this inequality
with the same constant.
Taking a positive sum or integral over accepted records and normalizing the nonzero accepted output proves Theorem~\ref{thm:fullrank}(a).
The Gaussian components from different records combine into a convex-Gaussian component,
whose weight lower bound remains $\eta_m(t)$.

\subsection{The sharp four-mode bound and white noise}
\label{subsec:noise-fourmode}
For the four-mode target $M$, the Gaussian-filter bound of
Appendix~\ref{app:spinor} further improves the estimate above to a sharp bound.
For four output modes, Lemma~\ref{lem:noise-compression} gives
$Y_K=B\cN_{t,k}(Z)B^\dagger$, with $k\le4$.
We first treat $k\le3$. It suffices to show that the compressed
$\cN_{t,k}(Z)$ lies in the convex-Gaussian cone:
the subsequent Gaussian map $B$ then keeps it in this cone,
giving target fidelity at most $1/2$.

For this purpose, a parity-resolved spectral decomposition of a physical positive operator
reduces the question to pure states of definite parity.
On three modes, an even pure state contains only vacuum and two-particle terms,
and hence is a BCS exponential whenever its vacuum amplitude is nonzero.
The case of zero vacuum amplitude follows from closedness of the set of Gaussian states.
Odd states can be converted to even states by Majorana parity conversion,
and fewer modes can be included by adding vacuum modes.
Thus, every physical positive operator on $k\le3$ modes lies in the
convex-Gaussian cone, completing the estimate for these branches.

For $k=4$, we use the attenuated target as a reference state, with target fidelity
\[
 a(t)=\Tr\!\left[P\cN_{t,4}(P)\right]
     =\frac{1+14t^4+t^8}{16}.
\]
Preparing the vacuum also attains fidelity $1/2$.
We will show that the larger of these two values,
\[
 h(t)=\max\left\{\frac12,a(t)\right\},
\]
is an upper bound on the output fidelity of every branch, namely
$\Tr(PY_K)\le h(t)\Tr Y_K$.
For fixed $B$, both sides are linear in $Z$,
so it suffices to prove the same upper bound for normalized pure preimages
of definite parity.
To apply the common Gaussian-filter bound of Appendix~\ref{app:spinor},
we first establish the following estimate for the spinor invariants.

\begin{lemma}[Four-mode attenuation bound]
\label{lem:noise-sharp-attenuation}
Let $\chi$ be a normalized four-mode vector of definite parity,
and let $X=\cN_{t,4}(\ket\chi\bra\chi)$, where $0\le t<1$. Then
\begin{equation}
 \max\{d_{\max}(X_+),d_{\max}(X_-)\}
 \le h(t)\,[A(X_+)+A(X_-)].
 \label{eq:noise-sharp-invariants}
\end{equation}
\end{lemma}

\begin{proof}
To calculate the invariants after attenuation, we first reduce the pure input
to a standard form containing only the vacuum and fully occupied terms.
Gaussian unitaries and Majorana conjugation commute with attenuation
and preserve the spinor singular values. Majorana conjugation exchanges the
two parity blocks and hence leaves the desired inequality unchanged.

First, we convert $\chi$ to the even sector by parity conversion.
In the real spinor coordinates of Appendix~\ref{app:spinor},
choose its overall phase so that $\chi^T\chi=\kappa\ge0$ and write $\chi=x+iy$.
Then $x\perp y$, with
$\|x\|^2=(1+\kappa)/2$ and $\|y\|^2=(1-\kappa)/2$.
The $\mathrm{SO}(8)$ action can rotate these two real directions to
those of $M$ and $i(\ket{0000}-\ket{1111})/\sqrt2$,
so in the occupation basis we may take
\[
 \ket{\chi_\kappa}
 =\alpha\ket{0000}+\beta\ket{1111},
 \qquad
 \alpha,\beta\ge0,\quad
 \alpha^2+\beta^2=1,\quad
 2\alpha\beta=\kappa.
\]

In this standard form, the attenuated matrix can be calculated directly.
Set $\mu=(1+t^2)/2$ and $\nu=(1-t^2)/2$.
The vacuum and fully occupied state are mapped to
$\operatorname{diag}(\mu,\nu)^{\otimes4}$ and
$\operatorname{diag}(\nu,\mu)^{\otimes4}$, respectively.
The operator $\ket{0000}\bra{1111}$ has Majorana degree four,
so its coefficient is multiplied by $t^4$.
Every configuration with $j$ occupied modes has the same diagonal entry
\[
 r_j=\alpha^2\mu^{4-j}\nu^j+\beta^2\mu^j\nu^{4-j},
 \qquad j=0,\ldots,4.
\]
Off-diagonal entries can occur only between the vacuum and fully occupied
configurations, with
$X_{0000,1111}=X_{1111,0000}=\kappa t^4/2$.

To use this matrix in the Gaussian-filter bound,
we need $A$ and $d_{\max}$ for its two parity blocks.
Let $\Theta$ be the conjugation associated with the invariant real
structure of Appendix~\ref{app:spinor}. In the occupation basis, it pairs
complementary configurations with fixed signs and exchanges the vacuum and fully occupied states.
The spinor singular values are the square roots of the eigenvalues of
$X_\pm(\Theta X_\pm\Theta^{-1})$,
so the calculation reduces to pairing these complementary configurations. Write
\[
 \delta=\mu^2\nu^2=r_2,\qquad
 D_0=\sqrt{r_0r_4},\qquad D_1=\sqrt{r_1r_3}.
\]
The six two-particle configurations in the even block give six
singular values $\delta$, while the two-dimensional vacuum and fully occupied part
gives $D_0\pm\kappa t^4/2$.
All eight singular values in the odd block equal $D_1$.
The definition of $r_j$ gives $D_0\ge\delta$,
and positivity of the two-dimensional part ensures $D_0\ge\kappa t^4/2$.
Hence
\begin{equation}
 \begin{aligned}
 d_{\max}(X_+)&=D_0+\frac{\kappa t^4}{2},
 & A(X_+)&=2D_0+6\delta,\\
 d_{\max}(X_-)&=D_1,
 & A(X_-)&=8D_1.
 \end{aligned}
 \label{eq:noise-sharp-singular-values}
\end{equation}

We now compare these invariants with the target case $\kappa=1$.
Let $A_{\rm tot}(\kappa)=A(X_+)+A(X_-)$.
For $\kappa=1$, $\chi_\kappa=M$, and the formulas above give
\[
 A_{\rm tot}(1)=1,\qquad
 d_{\max}(X_+(1))=a(t).
\]
To control the other values of $\kappa$, fix $t$. The definition of $r_j$ and
$4\alpha^2\beta^2=\kappa^2$ give
\[
 D_j(\kappa)^2-\kappa^2D_j(1)^2
 =(1-\kappa^2)\delta^2\ge0,\qquad j=0,1.
\]
Thus $D_j(\kappa)\ge\kappa D_j(1)$.
Write $h=h(t)$. Since $h\ge1/2$,
the coefficients of both $D_j$ in the difference $hA_{\rm tot}-d_{\max}(X_+)$ are nonnegative,
so these two lower bounds can be substituted directly:
\begin{equation}
 \begin{aligned}
 hA_{\rm tot}(\kappa)-d_{\max}(X_+)
 &=(2h-1)D_0(\kappa)+6h\delta
   +8hD_1(\kappa)-\frac{\kappa t^4}{2}\\
 &\ge6h\delta(1-\kappa)+\kappa[h-a(t)]
 \ge0.
 \end{aligned}
 \label{eq:noise-sharp-difference}
\end{equation}
For the odd block,
$d_{\max}(X_-)=A(X_-)/8\le hA_{\rm tot}$.
This proves Eq.~\eqref{eq:noise-sharp-invariants}.
\end{proof}

We return to an arbitrary physical preimage $Z$ and take its spectral decomposition
$Z=\sum_\ell\lambda_\ell\ket{\chi_\ell}\bra{\chi_\ell}$,
where $\lambda_\ell\ge0$ and the $\chi_\ell$ are normalized vectors
of definite parity.
Keep $B$ fixed and apply
Lemma~\ref{lem:noise-sharp-attenuation}
and Eq.~\eqref{eq:spinor-filter-bound} to each component,
then sum the resulting homogeneous inequalities with weights $\lambda_\ell$ to obtain
\begin{equation}
 \Tr(PY_K)\le h(t)\Tr Y_K.
 \label{eq:noise-fourmode-homogeneous}
\end{equation}
Since $h(t)\ge1/2$, this bound also covers the
$k\le3$ branches treated above.
The constant depends only on $t$. Taking a positive sum or an integral over accepted branches
and normalizing the nonzero accepted output proves
Theorem~\ref{thm:fullrank}(b).

This upper bound is attained for the source $\rho=\cN_{t,4}(P)$.
First, its smallest attenuation parameter is  $t$:
for any representation $\rho=\cN_{s,4}(\sigma)$, the parity expectation gives
\[
 t^8=s^8\Tr(\Pi_4\sigma)\le s^8.
\]
Thus, $s\ge t$, and $t$ itself is feasible, so $t_*(\rho)=t$.
Keeping one input copy attains fidelity $a(t)$,
and preparing the vacuum attains $1/2$.
Hence, $h(t)$ is attained for every $N\ge1$, which proves sharpness.
With $t=e^{-2\tau}$, this gives $t_*(\rho_\tau)=e^{-2\tau}$
and Corollary~\ref{cor:noise-models}(b).

For the white-noise source $\omega_F$ of Eq.~\eqref{eq:main-white-source},
where $F=1-15\epsilon/16$ and $0<\epsilon<1$,
it remains to determine its smallest attenuation parameter.
We begin again with parity: every representation
$\omega_F=\cN_{s,4}(\sigma)$ satisfies
\[
 1-\epsilon=s^8\Tr(\Pi_4\sigma)\le s^8,
\]
so $s\ge(1-\epsilon)^{1/8}$.
We next construct a physical preimage that attains this lower bound,
taking $t=(1-\epsilon)^{1/8}$.

To construct this preimage, first determine the Majorana degrees of $P$. Since the target $P$ has zero covariance, its degree-two coefficients vanish.
Moreover, $\Pi_4P=P$ makes the complementary degree-six coefficients vanish, giving
\begin{equation}
 P=\frac{I+\Pi_4}{16}+T_4,
 \qquad T_4\text{ has Majorana degree four}.
 \label{eq:noise-target-degrees}
\end{equation}
To attain the parity lower bound above, the preimage must lie in the even sector.
We seek it in the standard family $\rho_{F_0}$:
\[
 \rho_{F_0}
 =F_0P+\frac{1-F_0}{7}Q
 =\frac{I+\Pi_4}{16}+\frac{8F_0-1}{7}T_4.
\]
Attenuation multiplies $\Pi_4$ by $t^8$ and $T_4$ by $t^4$.
Since $t^8=1-\epsilon$ already matches the parity coefficient,
only the degree-four coefficient remains to be matched, requiring
\[
 t^4\frac{8F_0-1}{7}=1-\epsilon=t^8.
\]
This determines the required preimage:
\begin{equation}
 \begin{gathered}
 \omega_F=\cN_{t,4}(\rho_{F_0}),
 \qquad t=(1-\epsilon)^{1/8},\\
 F_0=\frac{1+7t^4}{8}
     =\frac{1+7\sqrt{1-\epsilon}}8\in(1/8,1).
 \end{gathered}
 \label{eq:noise-white-preimage}
\end{equation}
The range of $F_0$ ensures that $\rho_{F_0}$ is a physical state.
Thus, the parameter lower bound is attained, giving
\[
 t_*(\omega_F)=(1-\epsilon)^{1/8}.
\]
In the main-text range $0<\epsilon<8/11$, $h(t)=a(t)$.
Substituting this parameter into Eq.~\eqref{eq:noise-fourmode-homogeneous}
and using $\sqrt{1-\epsilon}\le1-\epsilon/2$
gives Corollary~\ref{cor:noise-models}(a).

\section{Odd-parity noise under pairing protocols}
\label{app:matching}
To prove Theorem~\ref{thm:matching}, we first establish the structure of four-mode outputs of a pairing network,
then use the propagation of single-Majorana errors to bound their largest eigenvalue,
and finally apply the Gaussian-filter bound of Appendix~\ref{app:spinor}.

The operations allowed are those of
Definition~\ref{def:pairing-protocols}. Pair creation gives pairing preparations; braids give signed Majorana
permutations; pair fusion gives bilinear measurements.
Choosing subsequent braids according to the measurement outcomes gives
classical feed-forward.
Dephasing and syndrome randomization also use random braid products,
with quartic stabilizers acting by unitary conjugation.
Combined with postselection and discarding modes, these operations form
a pairing network, and the FLO operation in the second stage can be chosen
as the identity.

In our occupation convention, $\ket{a_8}$ in Eq.~(13) of Ref.~\cite{Bravyi2006Universal} 
is  $\ket M$, and its isotropic input corresponds to $\rho_F$.
The convergence threshold is given in
Refs.~\cite{Bravyi2006Universal,Cwiklinski2026Purification}. This proves Proposition~\ref{prop:bravyi-containment}.
We use the same two-stage convention below.

Throughout, we retain $q=(1-p)/8$ and $P_a=\gamma_aP\gamma_a$.
Following the conventions of Appendix~\ref{app:spinor},
fix the classical choices, ancillary preparations, and measurement outcomes
of the pairing stage, and refine the discarded modes into occupation outcomes.
Write the unnormalized output of this complete record, before the final FLO,
as $X=X_+\oplus X_-$.
Its trace retains the actual probability of the record.
All spinor coordinates are those of Appendix~\ref{app:spinor}.

\subsection{The stabilizer structure of four-mode outputs}
\label{app:matching-real}

To apply Eq.~\eqref{eq:spinor-terminal},
we need to determine which outputs have real parity blocks.
A Majorana stabilizer state here is a pure state uniquely specified
by its simultaneous eigenvalues under a complete, independent, pairwise commuting
set of Hermitian even Majorana monomials.
Pairing states are the special case in which the generators can be chosen as disjoint bilinears.
When comparing different components, we compare only these monomials and their products,
allowing their eigenvalue signs to differ.

\begin{lemma}[Four-mode stabilizer states]
\label{lem:matching-stabilizer-structure}
Any four-mode pure Majorana stabilizer state whose stabilizer contains a bilinear
is a pairing state.
If its stabilizer contains no bilinear, an overall phase makes the state real
in the fixed spinor coordinates of its parity sector.
\end{lemma}

\begin{proof}
The expectation of a Majorana bilinear in a stabilizer state is either $0$ or $\pm1$.
If the bilinear belongs to the stabilizer monomial group,
it has a definite eigenvalue.
Otherwise, completeness of this group implies that it anticommutes with a stabilizer,
so its expectation vanishes.

If a bilinear stabilizer exists, the corresponding pair of Majoranas
fixes the occupation of one mode.
The remaining three modes are still in a pure state of definite parity.
By the three-mode result in Appendix~\ref{subsec:noise-fourmode},
this state is Gaussian, so the entire four-mode state is Gaussian as well.
For a pure Gaussian state, the purity condition gives
\[
 \sum_{b\ne a}
 \bigl|\langle i\gamma_a\gamma_b\rangle\bigr|^2=1.
\]
Since each two-point expectation is either $0$ or $\pm1$,
each Majorana has a nonzero two-point correlation with  one other Majorana.
These pairs are disjoint and each has a definite bilinear eigenvalue,
so the state is a pairing state.

If the stabilizer contains no bilinear, all two-point expectations vanish.
Using the pure-state normal form from the proof of
Lemma~\ref{lem:noise-sharp-attenuation},
a Gaussian unitary and, if necessary, a parity conversion
bring the state to
$\alpha\ket{0000}+\beta\ket{1111}$,
where $\alpha,\beta\ge0$ and $\alpha^2+\beta^2=1$.
Vanishing two-point expectations require $\alpha^2=\beta^2=1/2$,
so this normal form is $\ket M$.
Gaussian unitaries and Majorana parity conversion
act as real orthogonal maps in the fixed spinor coordinates,
up to an overall phase.
Since $\ket M$ is real, an overall phase also makes the original state real.
\end{proof}

We now apply this result to the pure components of $X$.
Under the standard convention associated with the occupation basis,
$\gamma_{2r-1}=a_r+a_r^\dagger$ and
$\gamma_{2r}=-i(a_r-a_r^\dagger)$.
Note that $\ket M$ is a simultaneous eigenstate of the four independent commuting even monomials
$\gamma_1\gamma_2\gamma_3\gamma_4$,
$\gamma_3\gamma_4\gamma_5\gamma_6$,
$\gamma_5\gamma_6\gamma_7\gamma_8$,
$\gamma_1\gamma_3\gamma_5\gamma_7$,
and is therefore a Majorana stabilizer state.
Above, $a_r$ is the annihilation operator of the $r$th occupation mode.
Applying any Majorana monomial changes only the stabilizer eigenvalue signs,
not the Majorana factors in the stabilizers.

Moreover, since the two-point expectations of $\ket M$ satisfy
$\langle M|\gamma_a\gamma_b|M\rangle=\delta_{ab}$,
the eight vectors $\gamma_a\ket M$ form an orthonormal basis of the odd sector. Hence,
\begin{equation}
 O=\sum_{a=1}^{8}P_a.
 \label{eq:match-odd-basis}
\end{equation}
It follows that $\rho_p^{\otimes N}$ is a positive sum of pure stabilizer projectors.
Each pure component is obtained from $\ket M^{\otimes N}$
by applying a Majorana monomial, so they share the same stabilizer monomials,
although their eigenvalues may differ.

Fixed pairing preparations and signed permutations preserve this property.
For a bilinear measurement,
if the measured operator already belongs to this set of monomials,
the specified outcome only determines which components survive.
Otherwise, replace one anticommuting generator by the measured bilinear,
and multiply the other anticommuting generators by the old generator being replaced.
This update depends only on commutation relations,
not on the previous eigenvalue signs.
After removing modes whose occupations have been fixed,
all surviving output components still share a complete set of commuting monomials.

If these output monomials contain a bilinear,
Lemma~\ref{lem:matching-stabilizer-structure}
shows that every component is a pairing state.
If they contain no bilinear, every component has a real projector
in the same fixed spinor coordinates.
Taking a positive sum of these projectors gives
\begin{equation}
 X\in\operatorname{cone}(\CG_4)
 \quad\text{or}\quad
 X_\pm=\overline{X_\pm}.
 \label{eq:match-real-alternative}
\end{equation}
The latter case is precisely the condition required by Eq.~\eqref{eq:spinor-terminal}.
It remains to bound the ratio of the largest eigenvalue of such an $X$ to its total trace.

\subsection{Path propagation and positive source replacement}
\label{app:matching-replacement}
\label{app:matching-directional}

Connecting the paired endpoints of the preparation, measurement, and permutation tensors
in a fixed record produces paths and closed loops.
The Majorana indices of the original sources remain endpoints of the paths;
the non-Gaussian source tensors themselves are not decomposed into pairing tensors.

\begin{lemma}[Path rule]
\label{lem:matching-path}
Fix a complete record of a pairing network.
Each output Majorana is connected along a path
to another output Majorana or to a Majorana of a source.

If $\gamma_j$ is connected to another output Majorana $\gamma_k$,
all nonzero pure components of the output
have the same eigenvalue of $i\gamma_j\gamma_k$.

If $\gamma_j$ is connected to $\gamma_a$ of a source,
then, with all other independent physical inputs held fixed,
replacing this source $\rho$ by $\gamma_a\rho\gamma_a$
replaces the unnormalized output $X$ by $\gamma_jX\gamma_j$.
\end{lemma}

\begin{proof}
For a fixed outcome, a measured mode that is subsequently reused
can be written as an occupation bra followed by the known occupation ket.
Thus, apart from the sources themselves, all tensors consist of
pairing preparations, pairing contractions, and signed permutations.
On a pairing tensor,
a Majorana insertion can be moved from one endpoint to its paired endpoint,
producing only a fixed phase of absolute value $1$.

Let $K$ be the Kraus map of the record,
and let $\widehat\gamma_a$ denote the Majorana corresponding to the selected source's local $\gamma_a$
in the global input mode ordering.
Transporting the insertion along the source--output path gives
\[
 K\widehat\gamma_a=e^{i\theta}\gamma_jK
\]
for a fixed phase $e^{i\theta}$.
This phase cancels in density-operator conjugation,
so conjugating the input by $\widehat\gamma_a$
changes the output to $\gamma_jX\gamma_j$.
The Jordan--Wigner parity strings in the global embedding
act only on the other input blocks.
Since these blocks are all physical,
the parity strings cancel under conjugation,
and the operation is  the stated local source replacement.

For an output--output path,
the same endpoint relation gives $(i\gamma_j\gamma_k)K=\pm K$,
with the sign determined by the fixed record.
Closed loops contribute only scalar factors.
If one such factor vanishes, the entire branch is zero,
and the conclusion still holds.
\end{proof}

To apply this propagation relation to the full source $\rho_p$,
we need to replace the remaining parts of the same source by allowed preparations.
For any fixed $a$,
Eq.~\eqref{eq:match-odd-basis} gives
\begin{equation}
 \rho_p=pP+qP_a+
 \frac q6
 \sum_{\substack{u<v\\u,v\ne a}}(P_u+P_v).
 \label{eq:match-positive-split}
\end{equation}
Each of the remaining seven projectors occurs six times in this sum.
Each equal pair also admits the decomposition
\begin{equation}
 P_u+P_v =\ket{g_{uv}^{+}}\bra{g_{uv}^{+}}
   +\ket{g_{uv}^{-}}\bra{g_{uv}^{-}},\qquad
 \ket{g_{uv}^{\pm}} =\frac{(\gamma_u\pm i\gamma_v)\ket M}{\sqrt2}.
 \label{eq:match-free-pair}
\end{equation}
These two normalized vectors are the two outputs obtained from $\gamma_u\ket M$
by measuring $i\gamma_u\gamma_v$.
They remain pure Majorana stabilizer states
and have definite eigenvalues of the measured bilinear.
By Lemma~\ref{lem:matching-stabilizer-structure},
both are pairing states.
Thus, the remaining terms of Eq.~\eqref{eq:match-positive-split}
can indeed be replaced by preparations allowed in the first stage.

Now take $1/2\le p\le1$.
We show that, for each output Majorana $\gamma_j$,
the same $X$ admits a decomposition
\begin{equation}
X=G_j+C_j,\qquad
 G_j\succeq0,\qquad C_j\in\operatorname{cone}(\CG_4),\qquad q\gamma_jG_j\gamma_j\preceq pG_j.
 \label{eq:match-directional}
\end{equation}
Here, $C_j$ collects the free contributions,
while $G_j$ is required only to be positive, not Gaussian.
The decomposition may differ for different $j$.

We proceed by induction on the number of original noisy sources.
With no noisy sources, all preparations are pairing states,
so $X\in\operatorname{cone}(\CG_4)$,
and we can take $G_j=0$ and $C_j=X$.
For a network containing noisy sources, we
apply Lemma~\ref{lem:matching-path} starting from $\gamma_j$.

If the path ends at another output,
the occupation of one mode is fixed.
By the three-mode result of Appendix~\ref{subsec:noise-fourmode},
the remaining physical state is convex-Gaussian.
Thus, the entire $X$ belongs to the convex-Gaussian cone,
and again we can take $G_j=0$ and $C_j=X$.

If the path ends at $\gamma_a$ of a noisy source,
we apply Eq.~\eqref{eq:match-positive-split} to that source,
keeping the other inputs and the record unchanged.
With the selected source replaced by $P$, denote the output by $Y_0\succeq0$.
The first two terms contribute
\begin{equation}
 G_j^{(0)}=pY_0+q\gamma_jY_0\gamma_j.
 \label{eq:match-noisy-pair}
\end{equation}
Since $p\ge q$, we have
\[
 pG_j^{(0)}-q\gamma_jG_j^{(0)}\gamma_j
 =(p^2-q^2)Y_0\succeq0.
\]
For each remaining term, Eq.~\eqref{eq:match-free-pair}
replaces this noisy source by a pairing preparation.
We apply the induction hypothesis to these terms, which contain one fewer noisy source,
then sum their decompositions with the original nonnegative coefficients,
including $G_j^{(0)}$ in $G_j$.
Positive sums preserve the required order inequality and the convex-Gaussian cone,
giving Eq.~\eqref{eq:match-directional}.

This induction always acts on unnormalized outputs,
so it retains the preparation coefficients and measurement weights of every term.
The source decomposition is used only in the proof;
no component labels are supplied to the protocol.

\subsection{From the spectral bound to the optimal fidelity}
\label{app:matching-bound}

Continue to assume $1/2\le p\le1$.
The zero output automatically satisfies the required homogeneous inequality,
so take $X\ne0$ below.
If $X\in\operatorname{cone}(\CG_4)$,
any final Gaussian processing can only yield outputs of fidelity at most $1/2$.

In the remaining case, Eq.~\eqref{eq:match-real-alternative}
ensures that both parity blocks are real.
Let $\lambda=\lambda_{\max}(X)$,
and choose a corresponding normalized real eigenvector $\phi$ of definite parity,
denoted by $s\in\{+,-\}$.
A Gaussian unitary and, if necessary, a parity conversion can map $\phi$ to $M$.
Applying Eq.~\eqref{eq:spinor-overlap}
to the block of $C_j$ in sector $s$ gives
\[
 2\langle\phi|C_j|\phi\rangle
 \le\Tr(\Pi_sC_j).
\]
Combining $X=G_j+C_j$ with
$\Tr(\Pi_sG_j)\ge\langle\phi|G_j|\phi\rangle$
yields
\[
 2\lambda-\Tr X_s
 \le\langle\phi|G_j|\phi\rangle.
\]
Conjugating the order inequality in Eq.~\eqref{eq:match-directional}
by $\gamma_j$ gives $qG_j\preceq p\gamma_jG_j\gamma_j$, and hence
\begin{equation}
 \begin{aligned}
 q\bigl(2\lambda-\Tr X_s\bigr)
 &\le q\langle\phi|G_j|\phi\rangle\\
 &\le p\langle\gamma_j\phi|G_j|\gamma_j\phi\rangle\\
 &\le p\langle\gamma_j\phi|X|\gamma_j\phi\rangle.
 \end{aligned}
 \label{eq:match-directional-witness}
\end{equation}

The real spinor $\phi$, like $M$, has vanishing two-point expectations.
Thus, the eight vectors $\gamma_j\phi$ form an orthonormal basis
of the opposite parity sector.
Summing Eq.~\eqref{eq:match-directional-witness} over $j$
and using $8q=1-p$ gives
\begin{equation}
 (1-p)\bigl(2\lambda-\Tr X_s\bigr) \le p\Tr X_{-s}, \quad p\Tr X-\lambda \ge(2p-1)\bigl(\Tr X_s-\lambda\bigr)\ge0.
 \label{eq:match-eight-budget}
\end{equation}
Therefore, $\lambda_{\max}(X)\le p\Tr X$.
In the case of real blocks,
Eq.~\eqref{eq:spinor-terminal} gives
$F_G^*(X)=\max\{1/2,\lambda_{\max}(X)/\Tr X\}\le p$.
Together with the convex-Gaussian case above,
every nonzero pairing record satisfies
\begin{equation}
 F_G^*(X)\le p,\qquad \frac12\le p\le1.
 \label{eq:match-highp}
\end{equation}
Here, we use $F_G^*$ as defined in Appendix~\ref{app:spinor},
so the conclusion already includes singular filters, parity-changing filters,
Gaussian ancillas, and general accepted Gaussian operations.

For $0\le p<1/2$, use
\begin{equation}
 \rho_p
 =2p\,\rho_{p=1/2}+(1-2p)\frac O8.
 \label{eq:match-lowp-expansion}
\end{equation}
The state $O/8$ is a uniform mixture of the eight odd occupation states,
each of which is a pairing preparation.
Keeping the original protocol and record fixed,
expand the finite tensor power of Eq.~\eqref{eq:match-lowp-expansion}.
Each term contains only copies of $\rho_{p=1/2}$ and pairing preparations,
so Eq.~\eqref{eq:match-highp} at $p=1/2$ applies to these terms.
Terms with no noisy sources are themselves free inputs
and likewise cannot exceed fidelity $1/2$.
Summing their unnormalized outputs with the original coefficients
gives the upper bound for the low-$p$ case.

Finally, different pairing records may use different terminal FLO operations,
but all corresponding outputs satisfy the same homogeneous inequality depending only on $p$.
Taking a positive sum or integral over accepted records gives
\begin{equation}
 \Tr(PY)\le\max\{1/2,p\}\Tr Y.
 \label{eq:match-final}
\end{equation}
Zero outputs are included.
Keeping one input copy attains fidelity $p$,
while discarding the inputs and preparing the vacuum attains $1/2$.
For every $N\ge1$, choosing whichever of these two operations has the higher fidelity
attains the upper bound, completing the proof of Theorem~\ref{thm:matching}.
\section{The twirled slice, Majorana noise, and parity information}
\label{app:geometry}

This appendix gives the Gaussian twirl used in
Eq.~\eqref{eq:triangle}, explains the Majorana-error origins of the two
noise vertices, and derives the poisoning trajectory underlying
Corollary~\ref{cor:noise-models}(b) and the correlated parity-flag example
discussed in Sec.~\ref{sec:discussion}.

\subsection{Twirl and free region}

We need an averaging operation that preserves target fidelity
while retaining only the total input weights in the subspaces associated with $P,Q,O$.
In the fixed real even spinor coordinates of Appendix~\ref{app:spinor},
$M=e_1$, and the subgroup of $\operatorname{Spin}(8)$ fixing this vector
is $\operatorname{Spin}(7)$.
Restricting the representations of the two parity sectors to this subgroup,
the even sector decomposes as $\mathbf1\oplus\mathbf7$,
corresponding to $P$ and $Q$, respectively.
The odd sector is its irreducible eight-dimensional spin representation
\cite{Baez2002Octonions}.

Take the normalized Haar average over the Gaussian unitaries in this subgroup.
By Schur's lemma,
the off-diagonal blocks between distinct irreducible subspaces vanish,
and the operator within each subspace becomes a multiple of the identity.
Thus, for any physical operator $X$,
the average gives
\begin{equation}
 \mathcal T(X)
 =\Tr(PX)P+\frac{\Tr(QX)}7Q+\frac{\Tr(OX)}8O.
 \label{eq:geometry-twirl}
\end{equation}
The total weights in all three subspaces are preserved,
with $\Tr(PX)$ being the unnormalized target weight.
For a normalized input,
the output belongs to the slice in Eq.~\eqref{eq:triangle},
with unchanged target fidelity.

This continuous group average can be implemented 
by a random mixture of finitely many Gaussian unitaries.
The conjugation maps above form a compact set in a finite-dimensional superoperator space,
and their convex hull is also compact, so the Haar average belongs to that convex hull.
Carath\'eodory's theorem then expresses it as a finite convex combination.
Selecting one of these Gaussian unitaries with the corresponding probability,
applying it, and discarding the selection record implements the same channel $\mathcal T$.

The even and odd blocks of $\rho(a,b,c)$ are
$aP+bQ/7$ and $cO/8$, respectively,
and both are real in the spinor coordinates above.
Their spectra are
$a,b/7,\ldots,b/7$ and $c/8,\ldots,c/8$, respectively.
We can therefore use the relations
$A(X)=\Tr X$ and $d_{\max}(X)=\lambda_{\max}(X)$ from Appendix~\ref{app:spinor}.
Substituting into Eq.~\eqref{eq:spinor-invariants} gives
\begin{equation}
 C\!\left(aP+\frac b7Q\right)=[a-b]_+,
 \qquad
 C\!\left(\frac c8O\right)=0.
 \label{eq:geometry-concurrence}
\end{equation}
The largest eigenvalue of the even block is $\max\{a,b/7\}$.
When this maximum is $b/7$, the concurrence vanishes,
so both cases are covered by $[a-b]_+$.

By the four-mode concurrence criterion
\cite{Oszmaniec2014Simulation},
$\rho(a,b,c)$ belongs to $\CG_4$ if and only if $a\le b$.
On the fixed-even-parity edge $c=0$,
normalization gives $a+b=1$,
so the free boundary is at $a=1/2$.
The odd-parity source $\rho_p$ corresponds to
$(a,b,c)=(p,0,1-p)$,
and its even block has concurrence $C(pP)=p$.
Thus, it is non-convex-Gaussian for every $p>0$.

\subsection{Majorana errors and attenuation}

We first identify the errors corresponding to $O/8$ and $Q/7$.
Selecting one of the eight Majoranas uniformly
and applying the corresponding single-Majorana error to the target
gives the output $\cD(P)/8$, where
\[
 \cD(X)=\sum_{j=1}^{8}\gamma_jX\gamma_j.
\]
Since $\ket M$ has zero covariance,
$\langle M|\gamma_j\gamma_k|M\rangle=\delta_{jk}$,
and the eight vectors $\gamma_j\ket M$
form an orthonormal basis of the odd sector.
Thus, $\cD(P)=O$,
and uniform single-Majorana errors give  $O/8$.

Applying two errors in succession gives the unnormalized output $\cD^2(P)$.
Each $\gamma_j$ exchanges the two parity sectors,
so $\gamma_jO\gamma_j=E$, and hence
$\cD^2(P)=\cD(O)=8E$.
In the expansion of $\cD^2$,
the eight terms that select the same Majorana twice cancel the error,
contributing $8P$ in total.
The remaining terms come from two distinct Majoranas,
with each unordered pair appearing twice.
Setting $U_{jk}=i\gamma_j\gamma_k$
gives the output of the uniform pair-error channel,
\begin{equation}
 \frac1{28}\sum_{j<k}U_{jk}PU_{jk}^{\dagger}
 =\frac{\cD^2(P)-8P}{56}
 =\frac Q7.
 \label{eq:geometry-pair-channel}
\end{equation}

The symmetric-poisoning model introduced in Sec.~\ref{sec:fullrank}
allows these single-Majorana errors to recur over time.
Specifically, let the eight errors occur through independent Poisson processes at equal rates,
and absorb the common rate into the dimensionless time $\tau$.
In units of $\tau$, each error then has rate $1$,
so the generator of the evolution is $\cD-8\,\mathrm{id}$.

Physical operators contain only even-degree Majorana monomials.
For such a monomial $\gamma_S$,
when $j\in S$, conjugation by $\gamma_j$ gives a minus sign, when $j\notin S$ it gives a plus sign.
Therefore,
\[
 (\cD-8\,\mathrm{id})(\gamma_S)
 =-2|S|\gamma_S.
\]
The evolution multiplies each degree-$|S|$ term by $e^{-2\tau|S|}$,
which is  attenuation with parameter $t=e^{-2\tau}$:
\begin{equation}
 e^{\tau(\cD-8\,\mathrm{id})}
 =\cN_{e^{-2\tau},4}
 \quad\text{on physical operators}.
 \label{eq:noise-semigroup}
\end{equation}

Now apply this evolution to the target,
writing $\rho_\tau=\cN_{e^{-2\tau},4}(P)$.
By Eq.~\eqref{eq:noise-target-degrees},
$P$ contains only degrees zero, four, and eight,
with $T_4$ denoting the degree-four part of that decomposition.
Attenuation multiplies these parts by $1,t^4,t^8$, respectively, so
\begin{equation}
\cN_{t,4}(P) =\frac{I+t^8\Pi_4}{16}+t^4T_4 =t^4P+\frac{(1-t^4)^2}{16}E +\frac{1-t^8}{16}O.
\label{eq:geometry-trajectory}
\end{equation}
Substituting $E=P+Q$ and $t=e^{-2\tau}$
gives the weights of $\rho_\tau$ in Eq.~\eqref{eq:triangle}:
\begin{equation}
 \begin{aligned}
 a(\tau)&=\frac{1+14e^{-8\tau}+e^{-16\tau}}{16},\\
 b(\tau)&=\frac{7(1-e^{-8\tau})^2}{16},\\
 c(\tau)&=\frac{1-e^{-16\tau}}2.
 \end{aligned}
 \label{eq:main-trajectory}
\end{equation}
Note that $a(\tau)=F_M(\rho_\tau)$ and,
in the weak-noise limit,
\[
 c(\tau)=8\tau+O(\tau^2),
 \qquad
 b(\tau)=28\tau^2+O(\tau^3).
\]
Thus, odd-parity noise appears at first order,
while even-parity noise orthogonal to the target appears at second order.
As the trajectory leaves $P$,
it is tangent to the odd-parity edge in Fig.~\ref{fig:landscape}.

For every $\tau>0$, $\rho_\tau$ has full rank.
Corollary~\ref{cor:noise-models}(b) gives its exact optimum
$\max\{1/2,a(\tau)\}$ for any number of input copies.
The optimal fidelity drops to $1/2$
at a different time from when the source enters the convex-Gaussian region.
Denote these times by $\tau_c$ and $\tau_*$, respectively,
and put $w=e^{-8\tau}$.
The condition $a(\tau)=1/2$ gives
$w^2+14w-7=0$,
whose positive root is $\sqrt{56}-7$.
The condition for entering the free region is $a=b$
in Eq.~\eqref{eq:geometry-concurrence},
or $3w^2-14w+3=0$,
whose root in $[0,1]$ is $(7-2\sqrt{10})/3$.
Therefore,
\[
 \tau_c =-\frac18\log(\sqrt{56}-7)
   \approx0.09089, \quad \tau_*=-\frac18\log\!\left(\frac{7-2\sqrt{10}}3\right)
   \approx0.18637.
\]
For $\tau_c\le\tau<\tau_*$,
the source remains non-convex-Gaussian,
but the optimal FLO target fidelity for every copy number
has already fallen to the value $1/2$ attainable by free preparation.

\subsection{Parity projection and supplied flags}
\label{subsec:parity-flags}

Reading the total parity of a source
and preserving coherence within a sector while doing so
are two distinct requirements.
An ideal nondestructive total-parity check
distinguishes only the even and odd outcomes,
with unnormalized branches
$X\mapsto EXE$ and $X\mapsto OXO$.
For $0<p<1$, its even branch satisfies
$E\rho_pE=pP$,
so it produces the target  with probability $p$.

This operation cannot be implemented by FLO.
Since $\rho_p$ has real parity blocks
in the spinor coordinates of Appendix~\ref{app:spinor},
Eq.~\eqref{eq:spinor-terminal} gives
\[
 F_G^*(\rho_p)=\max\{1/2,p\}<1.
\]
If this parity check could be implemented by FLO,
selecting the even outcome would give an output of fidelity $1$,
contradicting this single-input bound.
The excluded measurement is one that reads and selects a parity outcome
while preserving coherence within the selected sector.

Measurement of occupations mode by mode also determines total parity,
but simultaneously distinguishes different occupation configurations
within the same parity sector,
destroying coherence between those configurations.
Forgetting the detailed outcomes after the measurement
and retaining only their parity
cannot restore the lost coherence.
Thus, this readout is not equivalent to the nondestructive parity check above.

If source preparation already supplies a correlated record,
measuring the flag carrying that record can select a conditional state
without directly implementing the parity projection on the source.
Take $1/2<F<1$,
and let $\lambda\in(0,1)$ be the probability of a flag outcome $0$.
Place the flag mode last and consider the five-mode input
\begin{equation}
 \widehat\omega_{\lambda,F}
 =\lambda\rho_F\otimes\ket0\bra0
 +(1-\lambda)\frac O8\otimes\ket1\bra1.
 \label{eq:geometry-flag}
\end{equation}
The first term consists of an even source and an empty flag,
and the second of an odd source and an occupied flag,
so the entire input has definite even total five-mode parity.

Ignoring the flag gives the four-mode input
$\lambda\rho_F+(1-\lambda)O/8$,
whose weights in Eq.~\eqref{eq:triangle} are
\[
 (a,b,c)=\bigl(\lambda F,\lambda(1-F),1-\lambda\bigr).
\]
All three weights are strictly positive,
so this marginal has full rank on the entire four-mode Fock space.
Moreover,
$a-b=\lambda(2F-1)>0$,
so Eq.~\eqref{eq:geometry-concurrence}
shows that it remains non-convex-Gaussian.
Theorem~\ref{thm:fullrank}
therefore rules out distillation arbitrarily close to $\ket M$
using this marginal alone.

If the experimenter can read the supplied flag,
occupation outcome $0$ occurs with probability $\lambda$ and selects $\rho_F$.
This conditional state has fixed even parity and $F>1/2$,
so Theorem~\ref{thm:fixed-parity}
allows further distillation to arbitrary accuracy.
The target fidelity upon selection is $F$, ignoring the flag gives fidelity $\lambda F$.

This uses the source--flag correlation already present in the input.
The full-rank theorem constrains the four-mode marginal. An experimenter receiving the flag together with the source receives a different five-mode input.
An independently prepared free ancilla contains no information about source parity,
so measuring it cannot perform the same conditional selection.

\bibliography{references}
\end{document}